\documentclass[twocolumn,longbibliography,aps,prc,superscriptaddress,showpacs,floatfix]{revtex4-1}
\usepackage{graphics,epsfig,graphicx}
\usepackage{amsmath}
\usepackage{bm}
\usepackage{bbm}
\usepackage{amsfonts}
\usepackage{cancel}
\usepackage{multirow}
\usepackage[dvipsnames]{xcolor}
\newcommand{\abs}[1]{\left|#1\right|}
\newcommand{\vecr}{\bm{r}}
\newcommand{\vecp}{\bm{p}}
\newcommand{\veck}{\bm{k}}

\begin{document}

\title{The $nnn$ and $ppp$ correlation functions}

\author{A. Kievsky}
\affiliation{Istituto Nazionale di Fisica Nucleare, Largo Pontecorvo 3, 56127 Pisa, Italy}
\author{E. Garrido}
\affiliation{Instituto de Estructura de la Materia, IEM-CSIC, Serrano 123, E-28006 Madrid, Spain}
\author{M. Viviani}
\affiliation{Istituto Nazionale di Fisica Nucleare, Largo Pontecorvo 3, 56127 Pisa, Italy}
\author{L.E. Marcucci}
\affiliation{Physics Department, University of Pisa, Largo Pontecorvo 3, 56127 Pisa, Italy}
\affiliation{Istituto Nazionale di Fisica Nucleare, Largo Pontecorvo 3, 56127 Pisa, Italy}

\author{L. \v{S}erk\v{s}nyt\.{e}}
\affiliation{Physik Department E62, Technische Universit\"at M\"unchen, James-Franck-Straße 1, 85748  Garching, Germany}
\author{R. Del Grande}
\affiliation{Physik Department E62, Technische Universit\"at M\"unchen, James-Franck-Straße 1, 85748 Garching, Germany}
\affiliation{Excellence Cluster ORIGINS, Boltzmannstraße 2, 85748 Garching, Germany}

\begin{abstract}
Scattering experiments with three free nucleons in the ingoing channel are extremely challenging in terrestrial laboratories.
Recently, the ALICE Collaboration has successfully measured the scattering of three protons indirectly, by using the femtoscopy method 
in high-energy proton-proton collisions at the Large Hadron Collider. 
In order to establish a connection with current and future measurements of femtoscopic three-particle correlation functions, 
we analyse the scenarios involving $nnn$ and $ppp$ systems using the hyperspherical adiabatic basis.
The correlation function is a convolution of the source function and the corresponding scattering wave function. 
The finite size of the source allows for the use of the free scattering wave function in most of the adiabatic channels except the 
lowest ones. The scattering wave function has been computed using two different potential models: 
$(i)$ a spin-dependent Gaussian potential with parameters fixed to reproduce the scattering length and effective range and 
$(ii)$ the Argonne $v_{18}$ nucleon-nucleon interaction. Moreover, in the case of three protons, the Coulomb interaction has 
been considered in its hypercentral form. The results presented here have to be considered as a first step in the description 
of three-particle correlation functions using the hyperspherical adiabatic basis, opening the door to the investigation of other 
systems, such as the $pp\Lambda$ system. For completeness, the comparison with the measurement by the ALICE Collaboration is shown 
assuming different values of the source radius.
\end{abstract}


\maketitle

\section{Introduction}
In the past few years, the femtoscopy technique \cite{Wiedemann:1999qn,Heinz,fem2} has been applied in high-energy $pp$ and $p-$Pb collisions at the Large Hadron Collider (LHC) 
to study the residual strong interaction
between hadrons~\cite{lfabb2021}. 
In such collisions, particles are produced and emitted at 
relative distances 
of the order of a femtometer, in the range of the nuclear force. 
The effect of the mutual interaction between hadrons is reflected as a correlation signal in the momentum distributions of the detected particles which can be studied using correlation functions. The latter incorporate information on the emission process as well as on the final state interaction of the emitted pairs at the femtoscopic scale. 
Therefore, by measuring correlated particle pairs at low relative energies and comparing the yields to theoretical
predictions, it is possible to perform a new study of the hadron dynamics. 
The high precision measurements obtained by the ALICE Collaboration in the strangeness sector made it possible to test lattice calculations for the first time and to challenge effective field theory results (for a comprehensive review see~\cite{lfabb2021} and references therein). 

The next challenge is to extend the method to test the hadronic interactions in three-body systems. Recently, 
the $ppp$ and $pd$ correlation functions have been measured by the ALICE Collaboration~\cite{femtoppp, femtopd}. The interpretation of the former measurement requires a correct treatment of the three-proton scattering wave function which has to be used as input in the computation of the corresponding correlation function. This observable reflects a complex structure mainly determined when the three hadrons have low relative momenta. Traditional low-energy scattering experiments with three free hadrons in the ingoing channel are currently not yet available. Therefore the femtoscopic measurement gives a unique opportunity to study a $3\rightarrow 3$ scattering process. In the $pd$ case a very detailed discussion has been recently performed~\cite{pdtheory}, showing that the description of the data is possible when a very sophisticated $pd$ scattering wave function is used.

In the present work, we would like to set the basis for the study of the three-particle correlation function.
Specifically, we will focus on
the $ppp$ correlation function, whose description presents intrinsic difficulties
due to the long-range Coulomb interaction. The asymptotic description of three
charged particles has always attracted a lot of attention since it is present in
many different systems: atomic, nuclear and sub-nuclear systems (for a recent
discussion see Ref.~\cite{yakovlev}). In this preliminary study, we perform a
simplification in the asymptotic description of the three protons and
concentrate on the different steps needed in the computation of the $ppp$
correlation function. However, at the end of the study we give indications of the corrections due to a complete treatment of the Coulomb interaction.

We would like to stress that the present study represents the first step in the description of the three-particle correlation functions and will serve as a guideline for future studies of systems including hyperons such as the $pp\Lambda$ system.

The paper is organized as follows: after introducing the main ingredients of the method and the correlation function,  we briefly present the two-body case needed to define the notation and the procedure. The results are compared to previous calculations and the data published by the ALICE Collaboration~\cite{pSigma0}. Then we describe the three-body case, first the $nnn$ and then the $ppp$ system. In all cases, we discuss how to construct the free asymptotic wave function with the proper symmetry. In fact, more and more partial waves are needed to correctly  describe the correlation function when increasing the energy. The calculated $ppp$ correlation function is finally 
compared to measurements by the ALICE Collaboration in $pp$ collisions at center-of-mass energy 13 TeV at the LHC. The last section is devoted to the conclusions.

\section{Correlation function}

In order to compute the $ppp$ correlation function, two main ingredients are needed: the source function
and the $ppp$ wave function at different energies. 
The former is modelled as the product of three single Gaussian emitters, depending only on the size of the source. 
For the latter, we use the
hyperspherical harmonic (HH) method~\cite{HH2008,HH2020} in conjunction with the
adiabatic harmonic (HA) basis~\cite{garrido}. These two methods have been extensively used in the description of the three-nucleon continuum~\cite{kievsky2001,kievsky2004,gar14,higgins2020,higgins2021}.

In the present case, we apply these methods to describe the $ppp$ system. Essentially we decompose the
$ppp$ wave function in partial waves having well-defined values of total angular
momentum and parity, $J^\pi$. After introducing the hyperspherical coordinates
we solve the hyperangular Hamiltonian including the different partial waves compatible with the lowest values of the grand angular quantum number $K$. Here we show that the lowest adiabatic channel provides
sufficient accuracy. We pay particular attention to the construction of the different spatial and spin symmetries entering in the description of the wave function. We discuss these elements first in the $nnn$ case, where the Coulomb interaction is not present. Then we extend the analysis to the $ppp$ case considering the Coulomb interaction in the hypercentral approximation. In this way, the asymptotic wave function can
be evaluated analytically. It will be shown that the structure of the $ppp$ and the $nnn$ correlation functions are mostly determined by the lowest partial waves in which the nuclear interaction appreciably distorts the free scattering wave function. 

Regarding the $pp$ or $nn$ interactions, taking advantage of the fact that the $pp$ and $nn$ systems are located inside the universal window~\cite{kievsky2021}, we first use a  Gaussian potential model constructed to reproduce the corresponding scattering lengths and effective ranges. This potential acts only in $s$-wave. Then we incorporate in the study the results obtained using a more realistic potential, the Argonne $v_{18}$ potential (AV18)~\cite{AV18} already used in earlier studies of the $pp$ correlation function. At the end, we will estimate three-nucleon interaction effects using the AV18 potential in conjunction with the Urbana IX three-nucleon force~\cite{urbana}.

First, the two-body case will be presented. The study of the two-particle scattering through the measurement of their correlation
function is based on the femtoscopy method~\cite{Wiedemann:1999qn,fem2}. This method
is used in high-energy collisions and measures correlated pairs of particles having
low values of relative momentum. It was recently applied to measure hadron-hadron correlation functions such as $pp$~\cite{ppCF}, $pK^\pm$~\cite{Acharya:2019bsa,ALICE:2022yyh}, $p\mathrm{\Lambda}$~\cite{pLambda}, $p\mathrm{\Sigma}^0$~\cite{pSigma0}, $\mathrm{\Lambda}\mathrm{\Lambda}$~\cite{ALICE:LL}, $p\mathrm{\Xi}^-$~\cite{pXi}, $\Lambda K^\pm$~\cite{ALICE:2023wjz} $p\mathrm{\Omega}^-$~\cite{ALICE:pOmega}, $\Lambda\Xi$~\cite{ALICE:LambdaXi}, $p\phi$~\cite{ALICE:2021cpv} and baryon-antibaryon~\cite{ALICE:2021cyj}. The final state interaction between measured particles was then studied by comparing the experimental values to the
theoretical predictions. 

For two-body systems, the correlation function is defined in terms of the relative momentum $k$ between the particles in the pair rest frame by the Koonin-Pratt equation~\cite{Koonin,Pratt}
\begin{equation}
 C_{12}(k)=\int d\bm{r} \,  S_{12}(r) |\Psi_s|^2,
 \label{corr}
\end{equation}
where $S_{12}(r)$ is the so-called source function, which is an effective parameterization of the properties of the particle emission process, defined in terms of the relative distances $r$ between the particles, 
and $ |\Psi_s|^2$ is the square of the scattering wave function of the two particles. 

The correlation function can be generalised for a three-particle case as 
\begin{equation}
C_{123}(Q)=\int \rho^5 d\rho \,d\Omega_\rho \,S_{123}(\rho) |\Psi_s|^2,
\label{c123}
\end{equation}
where $S_{123}(\rho)$ is a three-particle source function. Here, $Q$ is the hypermomentum, $\rho$ is the hyperradius, $\Omega_\rho$ corresponds to the set of five hyperangles, and $|\Psi_s|^2$ is the square of the scattering wave function of the three particles. The details of these variables are given in Section~\ref{chap:threebodycase}. While the two-particle scattering wave function has been already estimated for many different hadron-hadron and hadron-nucleus pairs, it has never been evaluated for the three-particle case, which is the goal of this work.

\section{The two-body case}
As a preliminary step to analyse a three-particle scattering process,
we discuss first the scattering wave function for two protons. Considering only the Coulomb interaction, the scattering wave function for two protons is 
\begin{eqnarray}
\lefteqn{
	\Psi^0_s=
 } \label{eq1} \\ & & 
 4\pi\sum_{JJ_z}\sum_{\ell m S S_z} i^\ell \frac{F_\ell(\eta,kr)}{kr} (\ell m S S_z|JJ_z)
	 {\cal Y}^{JJ_z}_{\ell S}(\Omega_r) Y^*_{\ell m}(\Omega_k),
  \nonumber
\end{eqnarray}
where $\ell$, $S$, and $J$ are the relative orbital angular momentum, total spin, and total
angular momentum, respectively, with projections $m$, $S_z$, and $J_z$; and $\Omega_r$ and $\Omega_k$
are the polar and azimuthal angles describing the directions of the relative coordinate ($\bm{r}$) and the relative momentum ($\bm{k}$).

In the expression above, $F_\ell(\eta,kr)$ is the regular Coulomb function with 
Sommerfeld parameter $\eta=e^2 \mu/(\hbar^2k)$, where $\mu$ is the reduced mass, and 
\begin{equation}
	{\cal Y}_{\ell S}^{JJ_z}(\Omega_r)=\sum_{m S_z} (\ell m \, S S_z|JJ_z) Y_{\ell m}(\Omega_r) \chi_{SS_z} \,,
 \label{coup0}
\end{equation}
where $\chi_{SS_z}$ is the spin function arising from the coupling of two spin-$\frac{1}{2}$ particles
to $S$=0, 1, and ${Y}_{\ell m}(\Omega_r)$ is a spherical harmonic function. 

For two uncharged particles ($\eta=0$), the Coulomb function, $F_\ell(\eta,kr)$, reduces to the Riccati-Bessel
function $kr j_\ell(kr)$, and Eq.(\ref{eq1}) becomes:
\begin{equation}
\Psi_s^0=e^{i\bm{k}\cdot\bm{r}} \sum_{SS_z} \chi_{SS_z}.
\label{freewf}
\end{equation}

The expressions above are general, and valid for systems without any well-defined symmetry. However, for two identical nucleons the quantum numbers $\ell$
and $S$ are restricted to those combinations allowed from the antisymmetry requirement. In this case $\ell$ and $S$ are not independent, and we then 
introduce the index $[\ell S]$ indicating that $S$=0 ($S$=1) for even (odd) values of $\ell$.

The norm of the scattering wave function, $|\Psi^0_s|^2_\Omega$, is defined as the average over the angular coordinates of the square of the wave function, i.e., 
\begin{equation}
    |\Psi^0_s|^2_\Omega = \frac{1}{(4\pi)^2}\int d\Omega_r \int d\Omega_k |\Psi^0_s|^2,
    \label{norm}
\end{equation}
which for the free case (no Coulomb), without any symmetry, and after introducing Eq.~(\ref{freewf}), becomes 
\begin{equation}
|\Psi^0_s|^2_\Omega=N_S
\label{normfac}
\end{equation}
where $N_S=\displaystyle\sum_{SS_z} 1 =\sum_S(2S+1)=4$ is the number of spin states.

In the case of antisymmetric wave functions some structure appears, and use of Eq.(\ref{eq1}) leads to
\begin{equation}
|\Psi^0_s|^2_\Omega= \frac{2}{N_S} \sum_{[\ell S]} \left(\frac{F_\ell(\eta,kr)}{kr}\right)^2 N_{[\ell S]},
\end{equation}
where the factor of 2 is due to the fact that we are dealing with two identical
particles, and where the $1/N_S$ factor, see Eq.(\ref{normfac}), has been introduced to impose $|\Psi^0_s|^2_\Omega \rightarrow 1$ as $r \rightarrow\infty$.

The quantity $N_{[\ell S]}$ is the number of allowed states. Without considering a particular symmetry the number of states would be
$N_{[\ell S]}=4(2\ell +1)$. However, for antisymmetric states we have that
\begin{equation}
N_{[\ell S]}=\left\{ \begin{array}{cc}
             (2\ell +1) & \mbox{if $\ell$ even}  \\
             3 (2\ell +1) & \mbox{if $\ell$ odd} 
                     \end{array} \right.
\end{equation}
which leads to the following expression for the norm:
\begin{eqnarray}
|\Psi^0_s|^2_\Omega= \frac{1}{2}\sum_{\ell\equiv{\rm even}} \left(\frac{F_\ell(\eta,kr)}{kr}\right)^2 (2\ell +1) \nonumber \\
	+ \frac{3}{2}\sum_{\ell\equiv{\rm odd}} \left(\frac{F_\ell(\eta,kr)}{kr}\right)^2 (2\ell +1).
	\label{eq:asy0}
\end{eqnarray}

\begin{figure}[t]
\includegraphics[scale=0.44]{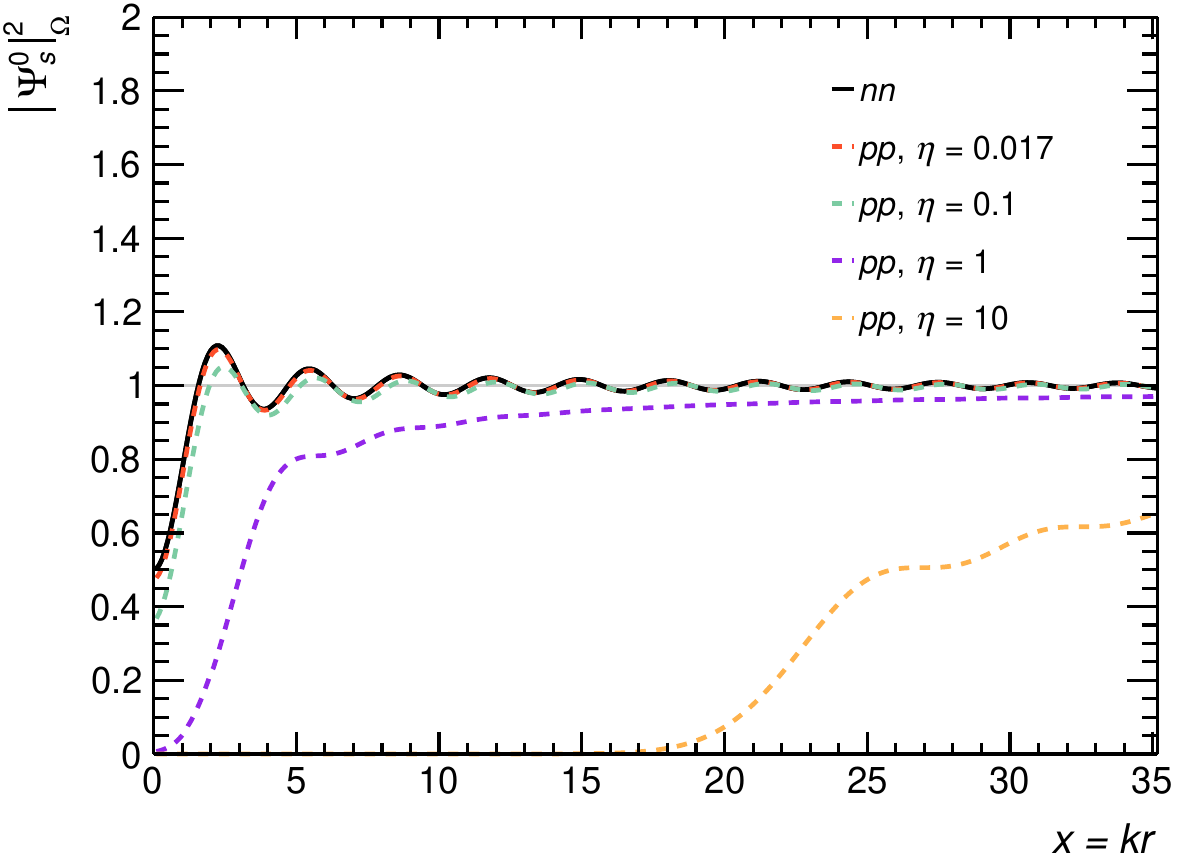}
\caption{The norm of the scattering wave function, Eq.(\ref{eq:asy0}), for the $nn$ and $pp$ cases. For the latter, different
	values of $\eta$ have been considered.}
\end{figure}

In Fig.1 the norm of the free scattering wave function, Eq.(\ref{eq:asy0}), is shown for the $nn$ and
$pp$ systems, considering different values of $\eta$ (for the $nn$ case, $\eta$=0 and $F_\ell(\eta, kr)$ has to be replaced
by $krj_\ell(kr)$). The convergence has been achieved after the inclusion of $\ell$-values up to $\ell=40$.

\subsection{Introducing the strong interaction}

For convenience let us write Eq.(\ref{eq1}) as:
\begin{equation}
	\Psi_s=4\pi\sum_{JJ_z}\sum_{\ell m S S_z} i^\ell \Psi_{\ell S}^{JJ_z} (\ell m S S_z|JJ_z) Y^*_{\ell m}(\Omega_k),
 \label{eq1b}
\end{equation}
where $\Psi_{\ell S}^{JJ_z}=(kr)^{-1}F_\ell(\eta,kr) {\cal Y}^{JJ_z}_{\ell S}(\Omega_r)$ is just the coordinate wave
function of the system with quantum numbers $\ell$, $S$, $J$, and $J_z$.

If the nuclear short-range interaction is considered,
the scattering wave function is still given by Eq.(\ref{eq1b}), but the coordinate wave function
takes now the form
\begin{equation}
    \Psi_{\ell S}^{JJ_z}= \sum_{\lambda S'} \frac{u^{\lambda S'}_{\ell S}\!(k,r)}{kr} {\cal Y}^{JJ_z}_{\lambda S'}(\hat r),
    \label{radeq}
\end{equation}
where we assume that, given an incoming channel with orbital angular momentum and spin $\{\ell, S\}$, the short-range interaction can mix it
with an outgoing channel with quantum numbers $\{\lambda, S'\}$.

The general large distance behaviour of the radial equations in Eq.(\ref{radeq}) is given by
\begin{equation}
    u^\lambda_{\ell}\rightarrow \delta_{\lambda \ell} F_\ell(\eta,kr)+T_{\lambda \ell} {\cal O}_\ell(\eta,kr),
\end{equation}
where for simplicity we have assumed that the potential is diagonal in the spin, 
$u_\ell^\lambda \equiv u_{\ell S}^{\lambda S}$, and 
${\cal O}_\ell(\eta,kr)=G_\ell(\eta,kr)+i F_\ell(\eta,kr)$ describes the outgoing wave function.
{\color{blue} 
In the expression above $T_{\lambda \ell}$ denotes the $T$-matrix elements which, 
in the case of a single channel, reduce to the usual form, 
$\sin \delta_\ell e^{i\delta_\ell}$, where $\delta_\ell$ is the phase shift.}

The antisymmetry requirement of the two-body wave function implies that $\{\ell,S\}$ and $\{\lambda, S\}$ become again $[\ell S]$ and $[\lambda S]$, 
which indicate that $S=0$ ($S=1$) for even (odd) values of $\ell$ and $\lambda$. Therefore, from Eqs.(\ref{eq1b}) and (\ref{radeq}), and following the same procedure as in the free case, the 
norm of the wave function results in
\begin{eqnarray}
|\Psi_s|^2_\Omega= \frac{1}{2}\sum_{\lambda, \ell\equiv{\rm even}} \left(\frac{u^\lambda_\ell(kr)}{kr}\right)^2 (2\ell +1) \nonumber \\
	+ \frac{3}{2}\sum_{\lambda,\ell\equiv{\rm odd}}
 \left(\frac{u^\lambda_\ell(kr)}{kr}\right)^2  (2\ell +1),
 \label{norm2}
\end{eqnarray}
 which trivially reduces to Eq.(\ref{eq:asy0}) when the short-range interaction is absent.

 \subsection{Integration on a spherical source}
 To model the correlation function in Eq.(\ref{corr}) a parametrization of the source function is required. Starting from a single particle emission source 
 of a Gaussian form, the following two-body source function is obtained 
 (Eq.(\ref{eq:twoParticleSource}) in the Appendix) 
\begin{equation}
 S_{12}(r)= \frac{1}{8\pi^{3/2}R^3} e^{-(r^2/4R^2)} \ ,
 \label{sour2b}
\end{equation}
 where $R$ is the source radius. The source function is normalised to unity in the coordinate space, and, therefore, it can be interpreted as the probability to emit particles at relative distance $r$. 

The ALICE Collaboration developed a data-driven approach, called the Resonance Source Model (RSM), able to describe the emission source in $pp$ collisions at the LHC \cite{ALICEsource}. The RSM assumes the existence of a common emission source for all baryons, composed of a Gaussian core, from which all primordial particles are emitted, and an exponential tail caused by the strong decays of resonances into particles of interest. 
In Ref.~\cite{ALICEsource}, the core radius was determined from the fit of the measured $pp$ correlation function as the $pp$ 
interaction is already well constrained from scattering and nuclear data and the corresponding wave function can be precisely 
calculated for different energies. The $pp$ correlation function was computed using the 
``Correlation Analysis Tool using the Schrödinger equation'' 
(CATS) \cite{CATS}, which is a numerical framework capable of evaluating the correlation function by taking as input either an 
interaction potential or a two-particle wave function, as well as an emission source of any form. The Schrödinger equation was 
solved for the AV18 potential including $s-$, $p-$ and $d-$waves, the Coulomb potential and properly antisymmetrising the $pp$ wave 
function. The fit was performed for different transverse mass ($m_T$) ranges of the pairs and the scaling trend as a function 
of $m_T$, typically observed in heavy-ion experiments, was found.  
{\color{blue}The transverse mass is defined as
$m_T = \left(k_T^{2} + m^{2}\right)^{1/2}$, 
where $k_T$ and $m$ are the
average transverse momentum and the average mass of the pair, respectively}.
Further, in Ref. \cite{ALICEsource} it is 
demonstrated that with the proper inclusion of the strongly decaying resonances, the obtained Gaussian core radius in 
$p\Lambda$ measurements is, indeed, identical to the $pp$ results, supporting the existence of a common emitting source 
for baryons in $pp$ collisions. 

The assumption of a common source for all the baryon-baryon pairs was used to test the interaction models of several hadron pairs and to access their low energy scattering properties through the correlation function. The source radius $R$ for the hadron pairs of interest is determined from the $m_T$ scaling, by using the average $m_T$ of the measured pairs and considering the effective enlargement induced by the strong decaying resonances. In the case of $pp$ pairs measured by the ALICE Collaboration, the effective $m_T$-integrated source radius amounts to $R= 1.249 \pm 0.008 \mathrm{(stat.)^{+0.024}_{-0.021}\mathrm{(syst.)}}$ fm \cite{pSigma0}.

As shown in the Appendix, Eq.~(\ref{eqa14}), since the source is spherical, the correlation function
can be computed as given in Eq.(\ref{corr}), but replacing $|\Psi_s|^2$ by
$|\Psi_s|^2_\Omega$. After insertion of Eq.({\ref{eq:asy0}}) the angular
integration can be trivially performed, and we obtain that for particles interacting only through the Coulomb force the correlation function is given by
\begin{eqnarray}
 & \displaystyle C^0_{12}(k)=\frac{1}{4\sqrt{\pi}R^3} \frac{1}{k^2} \int dr e^{-(r^2/4R^2)}  \times \\
 & \displaystyle \left(\sum_{\ell\equiv{\rm even}} F^2_\ell(\eta,kr) (2\ell +1) +
            {3}\sum_{\ell\equiv{\rm odd}} F^2_\ell(\eta,kr) (2\ell +1) \right). \nonumber
\end{eqnarray}

As an example, let us consider now the specific case in which the short-range interaction is limited 
to act on the $\ell=0$ singlet state. In this case the norm, Eq.(\ref{norm2}), can be written as
\begin{equation}
	|\Psi_s|^2_\Omega= |\Psi^0_s|^2_\Omega +\frac{1}{2} \left[\left(\frac{u_0(kr)}{(kr)}\right)^2- \left(\frac{F_0(\eta,kr)}{(kr)}\right)^2\right],
\end{equation}
where we have added and subtracted the $\ell=0$ free case contribution.
The correlation function results in
\begin{eqnarray}
\lefteqn{
	C_{12}(k)= C_{s}^0+C_{00} = }  \\ & &
=\int d\bm{r} S_{12}(r)\left[|\Psi^0_s|^2_\Omega- \frac{1}{2}\left(\frac{F_0(\eta,kr)}{kr}\right)^2+\frac{1}{2}\left(\frac{u_0(kr)}{kr}\right)^2 \right]
 \nonumber.
\end{eqnarray}

\begin{figure}[t]
\includegraphics[scale=0.44]{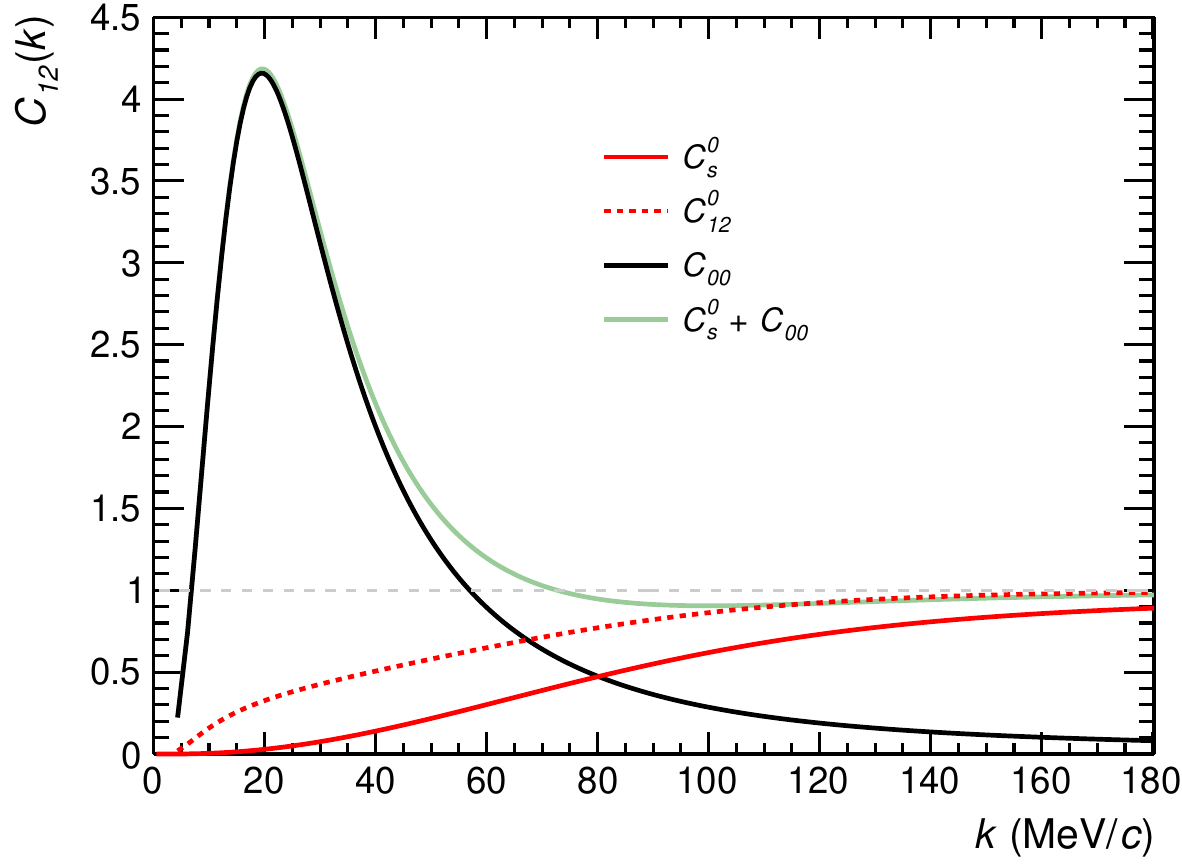}
\includegraphics[scale=0.44]{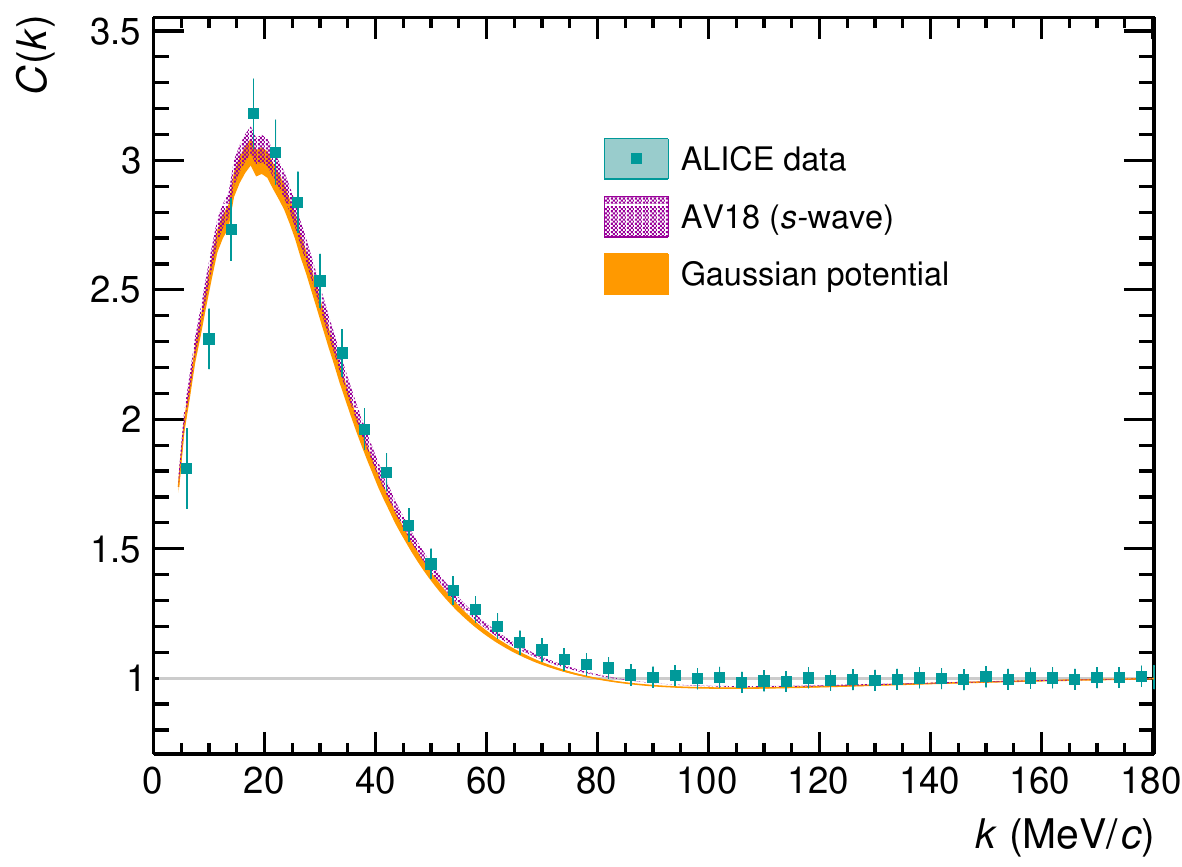}
    	\caption{{\it Top panel:} The $pp$ correlation function calculated using a Gaussian
	potential acting in the singlet $\ell=0$ channel. The source radius has been fixed to $R=1.249$ fm. The $\ell=0$ contribution is given by the black curve, the contribution of the other(free) partial
	waves is given by the red curve while the total result is given by the
	green curve. The correlation function considering only the Coulomb force
	is given by the red dashed curve.\\
{\it Bottom panel:} The $pp$ correlation function measured by the ALICE Collaboration (cyan data points) compared to the calculations obtained using the AV18 potential, considering only $s-$wave, (magenta band) and the prediction obtained using the Gaussian potential (orange band). The width of the bands represents the uncertainty due to the experimental determination of the source radius. The theoretical curves have been corrected to account for the experimental effects (see the text for the details).}
 \label{fig2}
\end{figure}
\begin{figure}
    \centering
    \label{fig:enter-label}
\end{figure}
The first term, $C_s^0$, is the component of the correlation function calculated
with the free wave function without including the $\ell=0$ contribution (the
first two terms of the integral in the equation above), whereas the second term,
$C_{00}$, is the $\ell=0$ component including the interaction. Both terms are
shown in Fig.~\ref{fig2} (top panel) using a source radius $R=1.249\,$fm
together with the correlation function, $C_{12}^0(k)$, considering only the Coulomb force. The $\ell=0$ radial wave function, $u_0(r)$, has been calculated using a Gaussian potential constructed to reproduce, in association with the Coulomb interaction, the $pp$ scattering length and effective range. Specifically
\begin{equation}
V_{pp}(r)=V_0 e^{-(r/r_0)^2} {\cal P}_0 + \frac{e^2}{r}
\label{eq:gausspp}
\end{equation}
with $V_0=-30.45\,$MeV and $r_0=1.815\,$fm, ${\cal P}_0$ is a projector on spin $S=0$.
In fact, the $s$-wave is well suited for a low-energy representation of the nucleon-nucleon (NN) potential 
by a two-parameter function, as the Gaussian used here, due to the dominance of the large value of the scattering
length. This introduces the two-nucleon system inside the universal window~\cite{kievsky2021} in which universal behaviour can be observed. A Gaussian representation of the NN interaction has been used many times in recent studies of the two-, three- and four-nucleon systems~\cite{higgins2020,higgins2021,tumino2023,gattobigio2019}, and even of nuclear matter~\cite{kievsky2018}. 
{\color{blue}}
The AV18 has been considered too, showing that essentially the two interactions give extremely close results for the $s$-wave component of the correlation function.

\subsection{Comparison with the experimental data}
\label{subsec:comparison2b}
In this Section, the computed $pp$ correlation function is compared to the experimental measurement published by the ALICE Collaboration~\cite{pSigma0}. The experimental correlation function is shown in Fig.~2 (bottom panel) with the cyan data points. The error bars include the combined contribution from the statistical and systematic uncertainties of the experiment, added in quadrature. 

The experimental correlation function is obtained in terms of the particle momentum distributions (the formal derivation is shown in Ref.~\cite{Wiedemann:1999qn})
\begin{equation}
    C(\bm{p}_1,\bm{p}_2) \equiv \frac{P(\bm{p}_1, \bm{p}_2)}{P(\bm{p}_1)\cdot P(\bm{p}_2)} \ ,
    \label{eq:CFdefinition1}
\end{equation}
where $P(\bm{p}_1, \bm{p}_2)$ is the two-particle momentum distribution, $P(\bm{p}_1)$ and $P(\bm{p}_2)$ are the single particle momentum distributions.
In absence of momentum correlations  $P(\bm{p}_1, \bm{p}_2) = P(\bm{p}_1)\cdot P(\bm{p}_2)$, leading to  
$C(\bm{p}_1,\bm{p}_2) = 1$.

After removal of the two-body center-of-mass motion, Eq.~\eqref{eq:CFdefinition1} turns to be
\begin{equation}
    C(k) = \mathcal{N}\frac{A(k)}{B(k)} \ ,
    \label{eq:CFdefinition2}
\end{equation}
where $k = |\bm{p}_1 - \bm{p}_2|/2$ is the relative momentum between the two protons, $A(k)$ and $B(k)$ are the relative momentum distribution of correlated and uncorrelated protons, respectively, and $\mathcal{N}$ is a normalisation constant.  In the experimental analyses, $B(k)$ is obtained by pairing particles emitted in different collisions, using the so-called event mixing technique~\cite{fem2}. The obtained mixed event distribution needs to be then properly normalised to 
$A(k)$ in a $k$ range where the interaction is absent.  This normalisation is absorbed in the factor $\mathcal{N}$. Following the arguments in Refs.~\cite{Wiedemann:1999qn,Pratt}, the correlation function in Eq.~\eqref{eq:CFdefinition2} can be related to the source and wave functions via the Koonin-Pratt formula introduced in 
Eq.~(\ref{corr}). 

To compare the experimental and the theoretical correlation functions, the latter has to be corrected for the following experimental effects: 1) momentum resolution of the detector which results in a smearing of the correlation function at low relative momenta; 2) the presence of secondary and misidentified protons in the experimental data sample. In the measurement performed by the ALICE Collaboration, the fraction of correctly identified primary protons amounts to $\lambda_{pp} = 0.67$, secondary protons are mainly produced in the decay of the $\Lambda$ hyperons, corresponding to a fraction $\lambda_{pp_\Lambda} = 0.203$, while the other secondary and misidentification contributions amount to $\lambda_\mathrm{X} = 0.127$~\cite{pSigma0}. The pairs with misidentified or secondary particles contribute to the measured correlation function as follows
\begin{equation}
    C (k) = \lambda_{pp} C_{pp} (k) + \lambda_{pp_\Lambda} C_{pp_\Lambda} (k) + \lambda_\mathrm{X} C_\mathrm{X} (k) \ .
\label{eq:2Bcorrection}
\end{equation}
The genuine $pp$ correlation function $C_{pp} (k)$ is obtained from the theoretical correlation function $C_{12} (k)$ shown in Fig. 2 top panel, further corrected for the momentum resolution of the detector. As shown in Eq. (\ref{eq:2Bcorrection}), the comparison with the experimental correlation function requires an additional scaling by the factor $\lambda_{pp}$. The contribution $C_{pp_\Lambda} (k)$ results from the pairing of primary protons with those emitted in the decay of a primary $\Lambda$ particle, carrying the effect of the primary $p\Lambda$ interaction.
The $p\Lambda$ correlation function has been modeled using chiral effective field theory calculations at the Next-To-Leading order
(NLO) \cite{NLO19} and transformed into
the relative momentum of the $pp$ pairs by applying the corresponding decay matrices \cite{pSigma0}. Following Ref.~\cite{pSigma0}, all the remaining contributions to the measured correlation function are assumed to be flat, i.e.\ $C_\mathrm{X}(k) = 1$.

The correlation function represented with a orange (magenta) band in the bottom panel of Fig.~2 is the resulting correlation function obtained from Eq.(\ref{eq:2Bcorrection}), by assuming the Gaussian (AV18) potential for the $pp$ $s$-wave interaction described in the previous section to calculate $C_{pp}(k)$. 
The width of the bands accounts for the experimental uncertainty on the source radius given in Ref.~\cite{pSigma0}. 
{\color{blue}In this reference the $pp$ correlation function was calculated using the AV18
interaction in $s$-, $p$- and $d-$waves showing that the effects of the $\ell >0$
partial waves are appreciable for $k > 100\,$MeV/c, very far from the peak at 20 MeV/c}.
The agreement of the calculations with the ALICE data provides the benchmark of the procedure that will be extended to the three-body sector in the next Section. 

\section{The three-body case} \label{chap:threebodycase}

Three-body wave functions are usually described by means of the Jacobi coordinates $\bm{x}$ and $\bm{y}$,
which for three identical particles read $\bm{x}=\bm{r}_2-\bm{r}_1$ and
$\bm{y}=\sqrt{4/3}\ [\bm{r}_3-(\bm{r}_1+\bm{r}_2)/2]$, where $\bm{r}_i$ is the position vector
of particle $i$. 

From these coordinates it is common to construct the hyperspherical coordinates,
which contain one radial coordinate, the hyperradius $\rho=(x^2+y^2)^{1/2}$, and
five hyperangles (the four angles describing the direction of $\bm{x}$ and $\bm{y}$ plus
$\alpha=\arctan(x/y)$) that we collect into $\Omega_\rho$. From the $\bm{x}$ and $\bm{y}$
conjugate momenta, $\bm{k}_x$ and $\bm{k}_y$, we can also construct the hypermomentum
$Q=(k_x^2+k_y^2)^{1/2}$ and the five hyperangles $\Omega_Q$ equivalent to $\Omega_\rho$,
but in momentum space.

These coordinates are the ones employed in the hyperspherical harmonic (HH) formalism 
(see Ref.~\cite{fabre1} and references therein). Within this method the three-body
scattering wave function can be written as described in detail in the appendix of 
Ref.~\cite{dan04}, and which takes the form:
\begin{eqnarray}
\lefteqn{
    \Psi_s=\frac{(2\pi)^3}{(Q\rho)^{5/2}}  \times } \label{3bdcon} \\ & &
    \sum_{JJ_z} \sum_{K\gamma}\Psi_{K\gamma}^{J J_z}
    \sum_{M_LM_S} (LM_L SM_S|JJ_z) {\cal Y}_{KL M_L}^{\ell_x\ell_y}(\Omega_Q)^*, \nonumber
\end{eqnarray}
where $\gamma$ groups the quantum numbers $\{\ell_x,\ell_y,L,s_x,S\}$. Then $\ell_x$ and $\ell_y$ are the relative orbital angular momenta associated to 
the Jacobi coordinates $\bm{x}$ and $\bm{y}$, which couple to the total orbital angular 
momentum $L$ (with projection $M_L$). The spin $s_x$ denotes the total spin of the two nucleons
connected by the $\bm{x}$ coordinate, which couples to the spin of the third nucleon to give
the total three-body spin $S$ (with projection $M_S$). The angular momenta $L$ and $S$ then
couple to the total angular momentum of the system $J$ with projection $J_z$.
Finally, $K=2\nu+\ell_x+\ell_y$ (with $\nu=0,1,2,\cdots$)  is the grand-angular momentum quantum number, and
${\cal Y}_{KLM_L}^{\ell_x \ell_y}$ are the usual hyperspherical harmonic functions.

Eq.~(\ref{3bdcon}) is the three-body partner of Eq.~(\ref{eq1b}). In fact, the coordinate
wave functions, $\Psi_{K\gamma}^{J J_z}$, take the general form
\begin{equation}
    \Psi_{K\gamma}^{J J_z}=\sum_{K'\gamma'} \Psi^{K'\gamma'}_{K\gamma}(Q,\rho) \Upsilon_{JJ_z}^{K'\gamma'}(\Omega_\rho)\ ,
    \label{3bdcoo}
\end{equation}
with
\begin{equation}
    \Upsilon_{JJ_z}^{K\gamma}(\Omega_\rho)=\sum_{M_L M_S} (L M_L S M_S |J J_z) 
    {\cal Y}_{KLM_L}^{\ell_x\ell_y}(\Omega_\rho) \chi_{SM_S}^{s_x},
    \label{upsilon}
\end{equation}
which are the three-body equivalent of Eq.(\ref{radeq}) and Eq.(\ref{coup0}), respectively. As in Eq.(\ref{radeq}), the radial wave function, $\Psi^{K'\gamma'}_{K\gamma}$, corresponds to 
a process with incoming and outgoing channels characterized by the set of quantum
numbers $\{K,\gamma\}$ and $\{K',\gamma'\}$, respectively.

Following Eq.(\ref{norm}), we define again the norm of the scattering wave function as the average over
the angular coordinates of the square of the wave function, i.e.:
\begin{equation}
    |\Psi_s|^2_\Omega = \frac{1}{\pi^6} \int d\Omega_\rho \int d\Omega_Q |\Psi_s|^2.
    \label{norm3b}
\end{equation}

The three-body wave function as defined above does not have a well-defined symmetry under
particle exchange. In order to introduce the correct symmetry we proceed as in the two-body case,
and consider in Eq.(\ref{upsilon}) only the HH functions that, coupled to the spin functions, provide the
correct total symmetry.

If the spin of the three-nucleon system is $S=\frac{1}{2}$, we have that Eq. (\ref{upsilon}) is given by 
\begin{eqnarray}
\lefteqn{
\Upsilon_{JJ_z}^{K\gamma }(S=\frac{1}{2})=} \label{eq:mixs} \\ && 
\sum_{M_L M_S} (L M_L \frac{1}{2} M_S |J J_z) \sum_\lambda (-1)^\lambda
    \frac{{\cal Y}_{KLM_L}^{\ell_x\ell_y,\bar{\lambda}}(\Omega_\rho) \chi^\lambda_{\frac{1}{2}M_S}}{\sqrt{2}}
	\nonumber
\end{eqnarray}
where we have introduced the HH functions ${\cal Y}_{KLM_L}^{\ell_x\ell_y,\bar{\lambda}}$ having well 
defined values of angular momentum $LM_L$
and mixed spin symmetry of type $\lambda$ coupled to the spin $S=\frac{1}{2}$ of three nucleons defined as
\begin{equation}
\chi^{\lambda}_{SS_z}=\sum_{\sigma_x\sigma_y} (\lambda \sigma_x \, \frac{1}{2} \sigma_y |SS_z) \chi_{\lambda \sigma_x} 
\chi_{\frac{1}{2}\sigma_y},
\end{equation}
where $\chi_{\lambda\sigma_x}$ and  $\chi_{\frac{1}{2}\sigma_y}$ are, respectively, the spin functions of the two-body system formed by the nucleons 1 and 2, and the one of the third nucleon. The quantum number $\lambda=1,0$ corresponds to the mixed spin
symmetry, symmetric or antisymmetric with respect to
the exchange of particles $1,2$, respectively. With $\bar\lambda$ we indicate
the conjugate symmetry, $\bar\lambda=1$ when $\lambda=0$ and vice versa. 

In the case of $S=\frac{3}{2}$, since the spin part is always symmetric under exchange of 
nucleons 1 and 2, we have that 
\begin{equation}
    \Upsilon_{JJ_z}^{K\gamma}(S=\frac{3}{2})=\sum_{M_L M_S} (L M_L \frac{3}{2} M_S |J J_z)
    {\cal Y}_{KLM_L}^{\ell_x\ell_y,a}(\Omega_\rho) \chi^1_{\frac{3}{2}M_S}
\end{equation}
where we have introduced the antisymmetric HH functions, ${\cal Y}_{KLM_L}^{\ell_x\ell_y,a}$, 
coupled to the symmetric spin $S=\frac{3}{2}$ of three nucleons. To be noticed that in both cases, $S=\frac{1}{2}$ and $\frac{3}{2}$, the index $\gamma$ does not include anymore the value of the spin of the two nucleons, $s_x$, since it is fixed by the symmetry requirements. This is similar to the
two-body case, where the value of $s_x$ is determined by the odd or even value
of $\ell_x$.

\subsection{The case of three free neutrons}

For the case of three free neutrons, since the Coulomb potential is absent, we have that \cite{gar14}
\begin{equation}
    \Psi^{K'\gamma'}_{K\gamma}(Q,\rho)=i^K \sqrt{Q\rho}J_{K+2}(Q\rho)\delta_{KK'}\delta{\gamma \gamma'},
    \label{prad}
\end{equation}
where $J_{K+2}(Q\rho)$ is the Bessel function of order $K+2$,
and the continuum wave function (\ref{3bdcon}) simply becomes
\begin{equation}
\Psi_s^0=e^{i\bm{Q}\cdot \bm{\rho}} \sum_{S M_s s_x} \chi_{SM_S}^{s_x}.
\label{free3b}
\end{equation}
Here the partial wave expansion of the three-body plane wave is given by:
\begin{eqnarray}
\lefteqn{
    e^{i\bm{Q}\cdot\bm{\rho}}=e^{i(\bm{k}_x\cdot \bm{x}+\bm{k}_y \cdot \bm{y} )}=} \label{3bdpw} \\ &&
    \frac{(2\pi)^3}{(Q\rho)^2}
\sum_{K\ell_x\ell_y L M_L} i^K J_{K+2}(Q\rho) 
{\cal Y}_{KL M_L}^{\ell_x\ell_y}(\Omega_\rho) {\cal Y}_{KL M_L}^{\ell_x\ell_y}(\Omega_Q)^*. \nonumber
\end{eqnarray}
Using this partial wave expansion, we can
verify that the norm of the three-body plane wave, Eq.(\ref{norm3b}), is given by 
\begin{equation}
\frac{16}{3(Q\rho)^4} \sum_{K} J_{K+2}^2(Q\rho)(K+3)(K+2)^2(K+1)=1,  \,
\end{equation}
where we have used the following property of the HH functions:
\begin{equation}
\sum_{\ell_x\ell_y L M_L} {\cal Y}_{KLM_L}^{\ell_x\ell_y*} {\cal Y}_{KLM_L}^{\ell_x\ell_y}=\frac{1}{12\pi^3}(K+3)(K+2)^2(K+1).
\label{eq:HHproperty}
\end{equation}

Insertion of Eq.(\ref{free3b}) into Eq.(\ref{norm3b}) leads to the same result for the norm of the free scattering function as in Eq.(\ref{normfac}), i.e.,
$|\Psi^0_s|^2_\Omega=N_S$, but where now $N_S=8$ (4 spin states from $S=\frac{3}{2}$, 2 from $S=\frac{1}{2}$ with $\lambda=0$, and 2 from $S=\frac{1}{2}$ with $\lambda=1$).

Considering antisymmetrization, we obtain that
the norm of the free three-neutron scattering state, Eq.(\ref{norm3b}), becomes
\begin{equation}
|\Psi^0_s|^2_\Omega = \frac{6}{N_S}\frac{2^6}{(Q\rho)^4} \sum_{K} J_{K+2}^2(Q\rho) N_{ST}(K)\ ,
\label{free3n}
\end{equation}
where, again, the factor of 6 ($=3!$) enters due to the fact that we are dealing
with three identical particles, the factor $1/N_S$ is introduced in order to
impose $|\Psi^0_s|^2_\Omega\rightarrow 1$ as $Q\rho\rightarrow \infty$,
and
where $N_{ST}(K)$ is the number of states, depending on the grand angular quantum number $K$.

To calculate $N_{ST}(K)$ we have to consider that for each value
of the grand angular quantum number, $K$, the HH functions can be symmetric, mixed, or antisymmetric.
For the three-nucleon system in isospin $T=3/2$, only the last two can contribute to the wave
function since the spin vector could 
be either of mixed symmetry ($S=\frac{1}{2}$) or symmetric ($S=\frac{3}{2}$). 
There is no antisymmetric spin state of three nucleons.
The two mixed symmetry spin states having $S=\frac{1}{2}$ and $\lambda=0,1$ combine with two mixed HH functions
resulting in an antisymmetric state, whereas the symmetric spin state with $S=\frac{3}{2}$ is combined with the
HH antisymmetric functions. Therefore the norm is
\begin{equation}
|\Psi^0_s|^2_\Omega = \frac{6}{N_S}\frac{2^6}{(Q\rho)^4} \sum_{K\ge1} J_{K+2}^2(Q\rho)( N^m_{ST}(K)+4N^a_{ST}(K))
\label{free3na}
\end{equation}
with $N_{ST}^m(K)$ the number of mixed HH functions and $N_{ST}^a(K)$ the number of antisymmetric HH functions
for each value of $K$. The factor $4$ in front of $N_{ST}^a(K)$ is the spin
degeneracy, whereas the spin degeneracy of $2$ in the case of the mixed symmetry
cancels out with the factor in the two-term sum as given in Eq.(\ref{eq:mixs}).
The fact that the spatially symmetric state is not present implies that the sum starts with $K=1$.

The following algorithm can be used to determine the number of HH functions having different symmetries. The number of HH functions
for a given $K$ is (see Eq.~(\ref{eq:HHproperty}))
\begin{equation}
    N=\frac{(K+1)(K+2)^2(K+3)}{12}.
    \label{eq:N}
\end{equation}
The ratio $r_n=N/N_{ST}^m$ results
\begin{equation}
    r_n = \left\{ \begin{array}{lr}
    \frac{3}{2}\frac{(n_3-1)(n_3+1)}{n_3^2+3} & \hspace{0.5cm} n_3-2=K \\
    \frac{1}{2}\frac{(n_3+1)(n_3+2)}{n_3(n+1)} & n_3-2 \neq K
    \end{array}
    \right.
\end{equation}
where we have introduced the integer $n$ defined as the integer part of $(K+2)/3$ and $n_3=3n$. Therefore $N_{ST}^m=N/r_n$. The number of antisymmetric HH functions is
\begin{equation}
    N_{ST}^a= \left\{ \begin{array}{lr}
    \frac{N-N_{ST}^m}{2} & \hspace{0.5cm} K\,\,{\rm odd} \\
    \frac{N-N_{ST}^m-m}{2} & K\,\, {\rm even}
    \end{array}
    \right.
\end{equation}
with $m=\frac{K}{2}+1$.
\begin{figure}[t]
\includegraphics[scale=0.44]{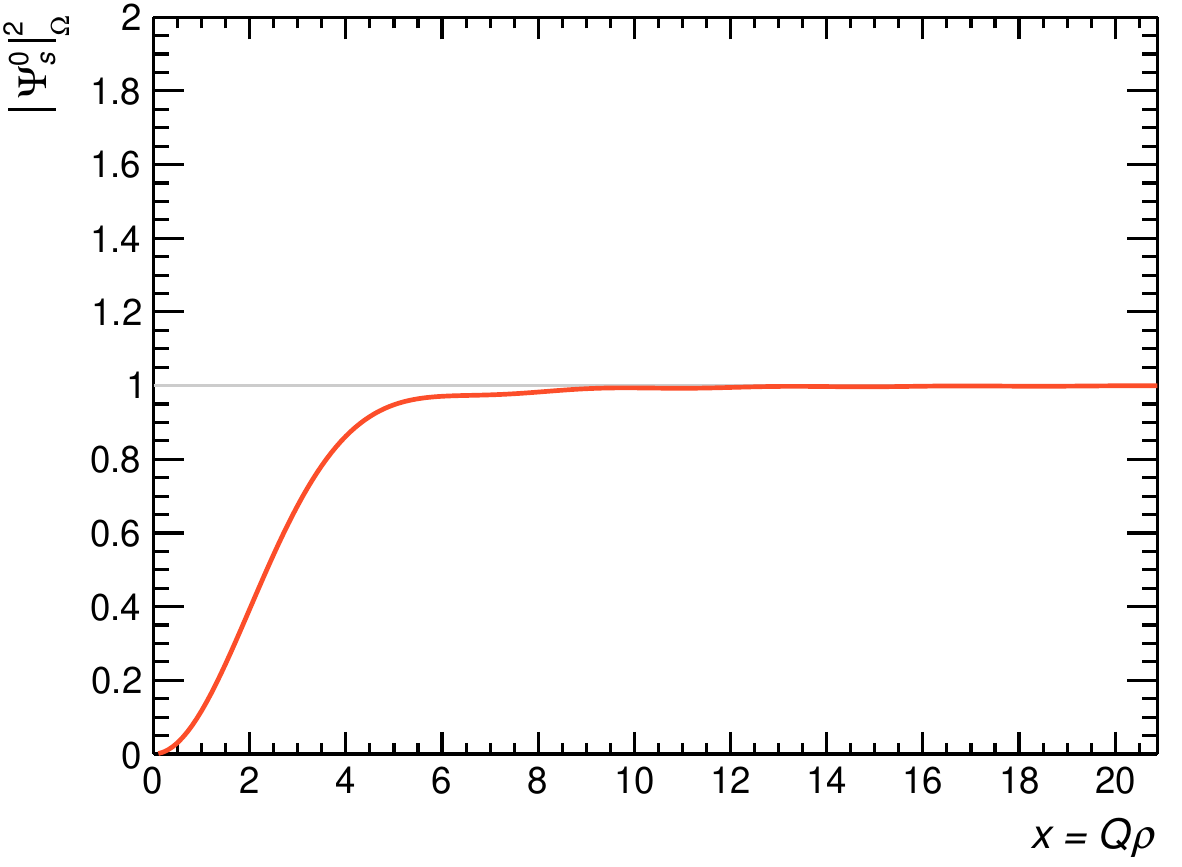}
	\caption{Norm of the free scattering wave function for three neutrons, Eq.(\ref{free3na}).}
 \label{fig3}
\end{figure}

With the above considerations, the result from Eq.(\ref{free3na}) is shown in Fig.~\ref{fig3}.  As seen in the figure, the norm tends to zero as 
$x \rightarrow 0 $, since the sum starts at $K=1$, and goes asymptotically to 1.

\subsection{Integration on a spherical source} \label{chap:3BCF}
In the case of three particles, the source function, analogous to Eq.(\ref{sour2b}), can be modelled as
\begin{equation}
S_{123}(\rho)=\frac{1}{\pi^3\rho_0^6}e^{-(\rho/\rho_0)^2}\ , \label{sour3b}
\end{equation}
with the normalization condition
\begin{equation}
\int S_{123}(\rho) \rho^5 d\rho \,d\Omega_\rho=1.
\end{equation}
The parameter $\rho_0$ of the three-particle source function is related to $R_M$, the parameter of the two-body source function, i.e. $\rho_0=2R_M$. This relation is discussed in  the Appendix. 
The value of $R_M$ must be determined experimentally from the $m_T$ distribution of proton pairs emitted in $ppp$ triplets. For this reason, we used a different notation with respect to the radius $R$ introduced in Section IIB. In general, the average $m_T$ of two protons which are emitted in pairs or triplets could be different in the experimental data sample and, additionally, the properties of the source function for three particles has not been studied yet. Recently, a numerical framework capable of simulating the effective emission source of a $n-$body system, based on the properties of the single particles, has been developed \cite{Mihaylov:2023pyl} but it has not been tested yet to the data due to the limitations in terms of statistics of the Run 2 ALICE data. This will be possible thanks to the larger statistics that has being acquired during the ongoing LHC Run 3 data campaign. By approximating $R_M \simeq R = 1.249$ fm, which is the value used to model the $pp$ correlation function, we find $\rho_0 \simeq 2.5$ fm. For the reasons mentioned above, this value of $\rho_0$ is not anchored to any realistic three-body source model and, therefore, a scan over different values of $\rho_0$ will be done in the next Sections.

The three-body correlation function is defined in Eq.~(\ref{c123}).
As in the two-body case (see the Appendix), due to the spherical symmetry
of the source function, we can replace $|\Psi_s|^2$, given in Eq.~(\ref{3bdcon}), 
by $|\Psi_s|^2_\Omega$, given in Eq.(\ref{norm3b}).
In the particular case of free $nnn$, making use of Eq.(\ref{free3n}) with $N_S=8$, we trivially
get from the definition in Eq.(\ref{c123}) that
\begin{equation}
C_{123}(Q)= \frac{6}{8}\frac{2^6}{Q^4\rho_0^6}
\int \rho\, d\rho\, e^{-\frac{\rho^2}{\rho_0^2}}\, \sum_K J^2_{K+2}(Q\rho) N_{ST}(K).
\label{cnnn}
\end{equation}

\begin{figure}[t]
\includegraphics[scale=0.44]{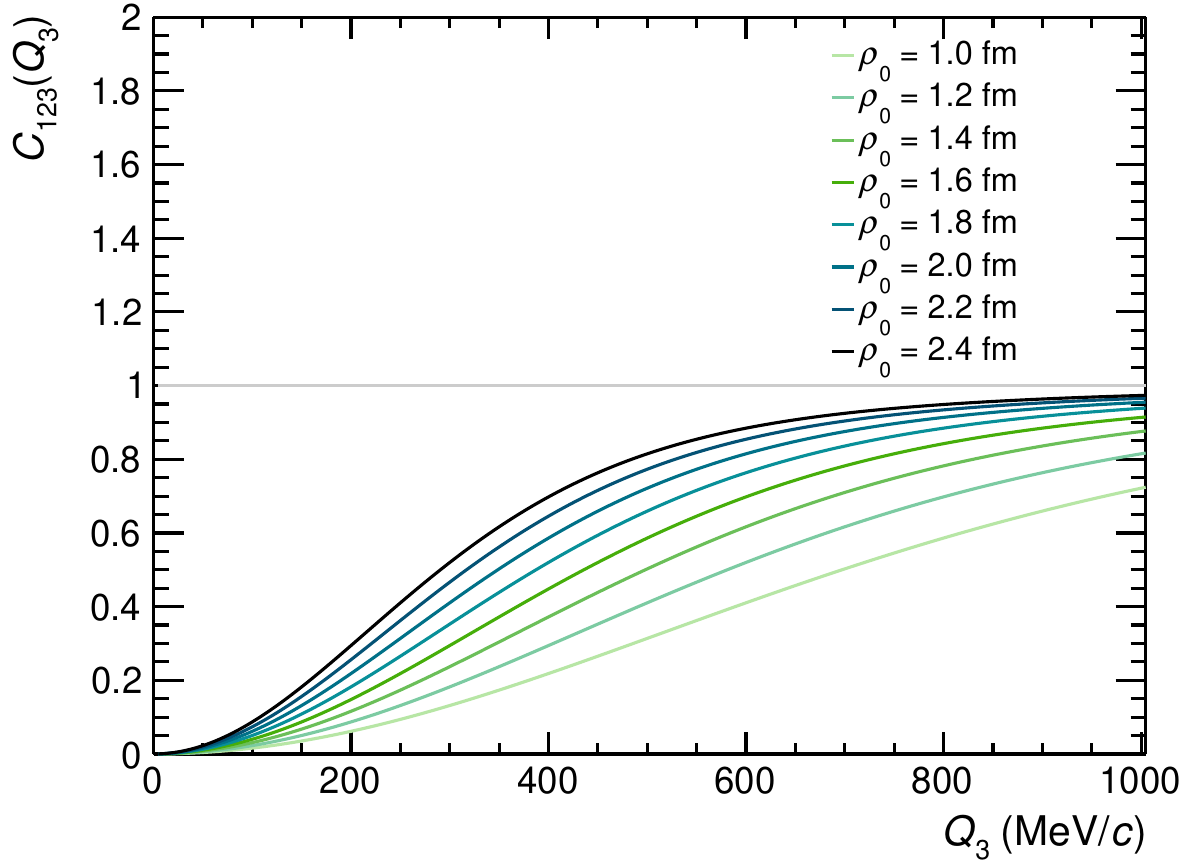}
	\caption{The correlation function using the free scattering wave function for three neutrons, Eq.(\ref{cnnn}),
	calculated using different source sizes.}
	\label{corr3n}
\end{figure}

In Fig.~\ref{corr3n} we can see the correlation function $C_{123}$ calculated using different source sizes. To be noticed that $Q$ refers to the total energy $E=\hbar^2Q^2/2m$ whereas the three-body momentum $Q_3$, used in some figures, is defined in Section~\ref{subsec:comparison}. It verifies $Q_3=\sqrt{6} Q$ and is the experimentally detected quantity.

\subsection{Introducing the interaction}

The analytical form of the three-body continuum wave function given in Eqs.(\ref{3bdcon}) and (\ref{3bdcoo}) is completely general.
As shown in Eq.(\ref{prad}), for three non-interacting nucleons, the matrix formed by the radial wave functions, $\Psi_{K\gamma}^{K'\gamma'}$,
is diagonal. However, when a short-range particle-particle interaction is present, this matrix is in general non-diagonal, in such a way
that the incoming and outgoing channels, characterized by quantum numbers $\{K,\gamma\}$ and
$\{K',\gamma'\}$, respectively, can be different.

When using the HH formalism, one of the difficulties is that the basis set used in the wave function expansion
in Eq.(\ref{3bdcoo}) can be quite large, which leads to a large system of coupled radial equations from which the matrix of radial wave functions can be obtained.

On many occasions, it is convenient to employ a different basis set, that we denote as $\Phi_n^{JJ_z}(\rho,\Omega_\rho)$,
where $n$ labels the different terms of the basis, depending not only on the hyperangles, $\Omega_\rho$,
but also on the hyperradius, $\rho$. The transformation between this basis set and the one in Eq.(\ref{upsilon}) is given
by:
\begin{equation}
    \Phi_n^{JJ_z}(\rho,\Omega_\rho) = \sum_{K\gamma} \langle \Upsilon_{JJ_z}^{K\gamma}(\Omega_\rho) | 
    \Phi_n^{JJ_z}(\rho,\Omega_\rho) \rangle_{\Omega_\rho}    \Upsilon_{JJ_z}^{K\gamma}(\Omega_\rho),
    \label{bastr}
\end{equation}
where the functions $\Upsilon_{JJ_z}^{K\gamma}(\Omega_\rho)$ are given in Eq.(\ref{upsilon}),
$\langle  |  \rangle_\Omega$ represents integration over the angular coordinates, and the summation 
over the HH quantum numbers $K$ and $\gamma\equiv\{\ell_x\ell_y L S \}$ has to be truncated at some maximum values
$K_\mathrm{max}$ and $\gamma_\mathrm{max}$.

Using the new basis set, the three-body continuum wave function can be written as described in detail in Appendix D
of Ref.~\cite{gar14}, and takes the form:
\begin{equation}
    \Psi_s = \frac{(2\pi)^3}{(Q\rho)^{5/2}} \sum_{JJ_z} \sum_n u_n^J 
    \sum_{s_x S M_s} \langle \chi_{SM_S}^{s_x} | \Phi_n^{JJ_z} (Q,\Omega_Q) \rangle^*,
\label{3bdad}
\end{equation}
where 
\begin{equation}
    u_n^J= \sum_{n'} u_n^{n'}(Q,\rho) \Phi_{n'}^{JJ_z}(\rho,\Omega_\rho),
    \label{3bdad2}
\end{equation}
and the incoming and outgoing channels are now $n$ and $n'$, respectively.

Eqs.(\ref{3bdad}) and (\ref{3bdad2}) are equivalent to Eqs. (\ref{3bdcon}) and (\ref{3bdcoo}) in the HH formalism. In fact, the latter 
are recovered from Eqs.(\ref{3bdad}) and (\ref{3bdad2}) simply by using that \cite{gar14}
\begin{eqnarray}
\lefteqn{
    u_n^{n'}(Q,\rho) =\sum_{K'\gamma'} \sum_{K\ell_x \ell_y L}  \Psi_{K\gamma}^{K'\gamma'}(Q,\rho) \times
    }  \\ &&
    \langle \Phi_{n'}^{JJ_z}(\rho,\Omega_\rho)|\Upsilon_{JJ_z}^{K'\gamma'}(\Omega_\rho)\rangle_{\Omega_\rho}
    \langle \Upsilon_{JJ_z}^{K\gamma}(\Omega_Q) | \Phi_{n}^{JJ_z}(Q,\Omega_Q) \rangle_{\Omega_Q}, \nonumber
\end{eqnarray}
and recalling that $\sum_n |\Phi_n\rangle \langle \Phi_n |=\mathbbm{1}$.

In this work, we shall use the hyperspherical adiabatic (HA) expansion method, described in detail in Ref.~\cite{garrido},
where the $\{ \Phi_n\}$ basis set is chosen to be formed by the eigenfunctions of the angular part of the Hamiltonian
equation:
\begin{equation}
    H_\Omega \Phi^{J J_z}_n(\rho,\Omega_\rho)=U_n(\rho) \Phi^{J J_z}_n(\rho,\Omega_\rho),
    \label{adiab}
\end{equation}
in such a way that the eigenvalue functions, $U_n(\rho)$, enter as effective potentials in a coupled
set of radial equations from which the $u_n^{n'}$ radial functions are obtained (see \cite{garrido} for 
details).

One of the main advantages of the HA method is that most of the dynamics of the system is captured by the
lowest terms in the adiabatic expansion in Eq.~(\ref{3bdad}). As a consequence, very few terms in the expansion, typically around 10, are enough
for an accurate description of the continuum wave function~\cite{gar14,gar12}. 
This reduces drastically the number of radial equations to be computed, and therefore the size of the matrix
formed by the radial wave functions, $u_n^{n'}$.

The reason for such a convenient behaviour of the HA expansion can be understood by noting that when only short-range interactions are involved (like in the $nnn$ case), each
adiabatic potential, $U_n(\rho)$ in Eq.(\ref{adiab}) behaves asymptotically as $K(K+4)/\rho^2$. In such a way,
each term of the HA basis tends to a single HH basis element term:
 \begin{equation}
	\Phi^{J J_z}_n(\rho,\Omega)\rightarrow \Upsilon_{JJ_z}^{K\gamma}(\Omega)\ .
 \label{hhad}
\end{equation}
This permits to associate a specific grand-angular momentum value, $K$,
to each adiabatic channel $n$. In fact, the asymptotic behaviour of the continuum
$u_n^{n'}(Q,\rho)$ functions takes the form:
\begin{eqnarray}
\lefteqn{
	u_n^{n'}(Q,\rho\rightarrow\infty) \rightarrow 
 } \label{asymu} \\ & &
		i^{K'}\sqrt{Q\rho} \left[ \delta_{KK'}J_{K'+2}(Q\rho)+ T_{KK'} {\cal O}_{K'+2}(Q\rho) \right],
  \nonumber
\end{eqnarray}
where $K$ and $K'$ are the grand-angular momentum values associated to the incoming and
outgoing channels $n$ and $n'$, respectively, ${\cal O}_{K+2}(Q\rho)=Y_{K+2}(Q\rho)+i J_{K+2}(Q\rho)$ is the 
the outgoing asymptotic wave function,
$T_{KK'}$ is a $T$-matrix element, and $J_{K+2}(Q\rho),Y_{K+2}(Q\rho)$ are the regular and 
irregular Bessel functions. 

It should be noticed that when inserting Eq.(\ref{asymu}) into the expression of the full
continuum wave function, Eq.(\ref{3bdad}), the second term, proportional to the $T$-matrix, tends to
zero as $K$ increases, since the interaction that produces the coupling between the different
HH channels remains hidden by the centrifugal barrier $K(K+4)/\rho^2$. On the other hand
channels with increasing values of $K$ are important as the energy increases.

Following Eqs.(\ref{free3n}) and (\ref{asymu}), the norm of the scattering wave function for the three-body
state with total angular momentum $J$, tends asymptotically to
\begin{eqnarray}
\lefteqn{
	|\Psi^{J}|^2_\Omega(\rho \rightarrow \infty)= 
}  \\ & &
 (2J+1) \frac{6}{8}\frac{2^6}{(Q\rho)^5}\sum_{KK'}
	|u_K^{K'}(Q,\rho\rightarrow\infty)|^2 N_{ST}(K),
 \nonumber
\end{eqnarray}
where $u_K^{K'}(Q,\rho\rightarrow\infty)$ is normalized as given in Eq.(\ref{asymu}), and the $(2J+1)$ factors appears after
summation over all the possible $J_z$ projection quantum numbers.
After summation over all the possible $J$ states, the norm becomes
\begin{eqnarray}
\lefteqn{
	\sum_{J} |\Psi^{J}|^2_\Omega= \frac{6}{8}\frac{2^6}{(Q\rho)^4} 
 \sum_J (2J+1)\times
 }    \label{eq:normf} \\ & &
  \left( \sum_{KK'}^{K_0} 
	\left|\frac{u_ K^{K'}(Q,\rho)}{\sqrt{Q\rho}}\right|^2
	+ \sum_{K>K_0} J_{K+2}^2(Q\rho) \right) N_{ST}(K) \ ,
	\nonumber
\end{eqnarray}
where $K_0$ is the quantum number indicating the maximum value of $K$ at which the interaction distorts
the free scattering state. For $K>K_0$ the wave function is taken as the free solution, i.e., for $K>K_0$,  $u_K^{K'}(Q,\rho)$ is replaced by
$i^{K'}\sqrt{Q\rho} \delta_{KK'} J_{K'+2}(Q\rho)$.

\subsection{The $nnn$ correlation function}

In order to calculate the three-neutron correlation function, we 
model the $nn$ interaction with a Gaussian potential for the singlet channel, $S=0$,
\begin{equation}
	V_{nn}(r)=V_0 e^{-(r/r_0)^2} {\cal P}_0\ ,
 \label{vnn}
\end{equation}
with the parameters $V_0=-30.42$ MeV and $r_0=1.8148$ fm selected to reproduce the $nn$ $s$-wave
scattering length and effective range of $-18.9\pm 0.4\,$fm and $2.8\pm0.1\,$fm, respectively ~\cite{machleidt2011}.

Since the spatially symmetric state is not present in the $nnn$ system, we have that 
the only states for which the lowest adiabatic HH channel (going asymptotically to $K=1$) contributes 
are the  $J^\pi=1/2^-,3/2^-$ states.

We first consider the $J^\pi=1/2^-$ state with total angular momentum $L=1$ ($\ell_x=0$, $\ell_y=1$) and total spin $S=1/2$.
If we consider the $K=1$ adiabatic channel only, making use of Eqs.(\ref{3bdad2}), (\ref{hhad}), and (\ref{asymu}) we get that the asymptotic form of Eq.(\ref{3bdad}) is given by
\begin{eqnarray}
\lefteqn{
	\Psi^{1/2^-}_{\rm HA} \stackrel{\rho \rightarrow \infty}{\longrightarrow}
 } \label{Psihas} \\ &&
	\frac{i (2\pi)^3}{(Q\rho)^{2}}
	{[}J_3(Q\rho)+T_{11}{\cal O}_3(Q\rho){]} \Upsilon_{\frac{1}{2}^-}^{1\gamma_1}(\Omega_\rho)
   \langle \chi_{S=\frac{1}{2}} | \Upsilon_{\frac{1}{2}^-}^{1\gamma_1}(\Omega_Q)\rangle ,
 \nonumber
\end{eqnarray}
where $\gamma_1\equiv \{ \ell_x=0, \ell_y=1, L=1, S=\frac{1}{2} \}$.

If we assume that the interaction distorts very little the behavior of the adiabatic channels
for $n>1$, the wave function of the $1/2^-$ state can be written as
\begin{equation}
	\Psi^{1/2^-}=\Psi^{1/2^-}_{\rm HA} + \Psi^{1/2^-}_0 \ ,
	\label{eq:psi1}
\end{equation}
where the free wave function is defined as follows
\begin{eqnarray}
\lefteqn{
	\Psi^{1/2^-}_0= 
 } \label{p12wf} \\ & &
 \frac{(2\pi)^3}{(Q\rho)^2}\sum_{K\ge 3} i^K 
	J_{K+2}(Q\rho) 
 \Upsilon_{\frac{1}{2}^-}^{K\gamma_1}(\Omega_\rho)
   \langle \chi_{S=\frac{1}{2}} | \Upsilon_{\frac{1}{2}^-}^{K\gamma_1}(\Omega_Q)\rangle .
\nonumber
\end{eqnarray}
Here the sum over $K$ starts at $K=3$ since the $K=1$ asymptotic term is included in 
$\Psi^{1/2^-}_{\rm HA}$. 

The norm of the wave function of Eq.(\ref{eq:psi1}) can be computed as given in Eq.(\ref{norm3b}),
and it can be written as:
\begin{eqnarray}
\lefteqn{\hspace*{-1cm}
|\Psi^{1/2^-}|^2_\Omega=
|\Psi_\mathrm{HA}^{1/2^-}|^2_\Omega+|\Psi_0^{1/2^-}|^2_\Omega + 
} \nonumber \\ & &
\langle \Psi_\mathrm{HA}^{1/2^-} | \Psi_0^{1/2^-} \rangle_{\Omega}+
\langle \Psi_0^{1/2^-} | \Psi_\mathrm{HA}^{1/2^-} \rangle_{\Omega},
\label{fullnorm}
\end{eqnarray}
where the first two terms in the r.h.s. are the norm of $\Psi_{\rm HA}^{1/2^-}$
and $\Psi_0^{1/2^-}$, respectively. The last two terms arise from the
interference between the two wave functions, and, by use of Eq.(\ref{3bdad}) and
Eq.(\ref{p12wf}), one can easily get that
\begin{equation}
	\langle\Psi^{1/2^-}_{\rm HA}|\Psi^{1/2^-}_0\rangle_{\Omega}=\frac{(2\pi)^6}{(Q\rho)^{5}}
 \sum_{K>1}^{K_\mathrm{max}}w^K_1(\rho)  J_{K+2}(Q\rho),
\label{eq:int}
\end{equation}
where $w_1^K(\rho)=\sqrt{Q\rho}\, u_1^1(\rho) \langle \Upsilon_{\frac{1}{2}^-}^{K\gamma} | \Phi_1^{\frac{1}{2}^-}\rangle_{\Omega_\rho}$.

\begin{figure}[t]
\includegraphics[scale=0.44]{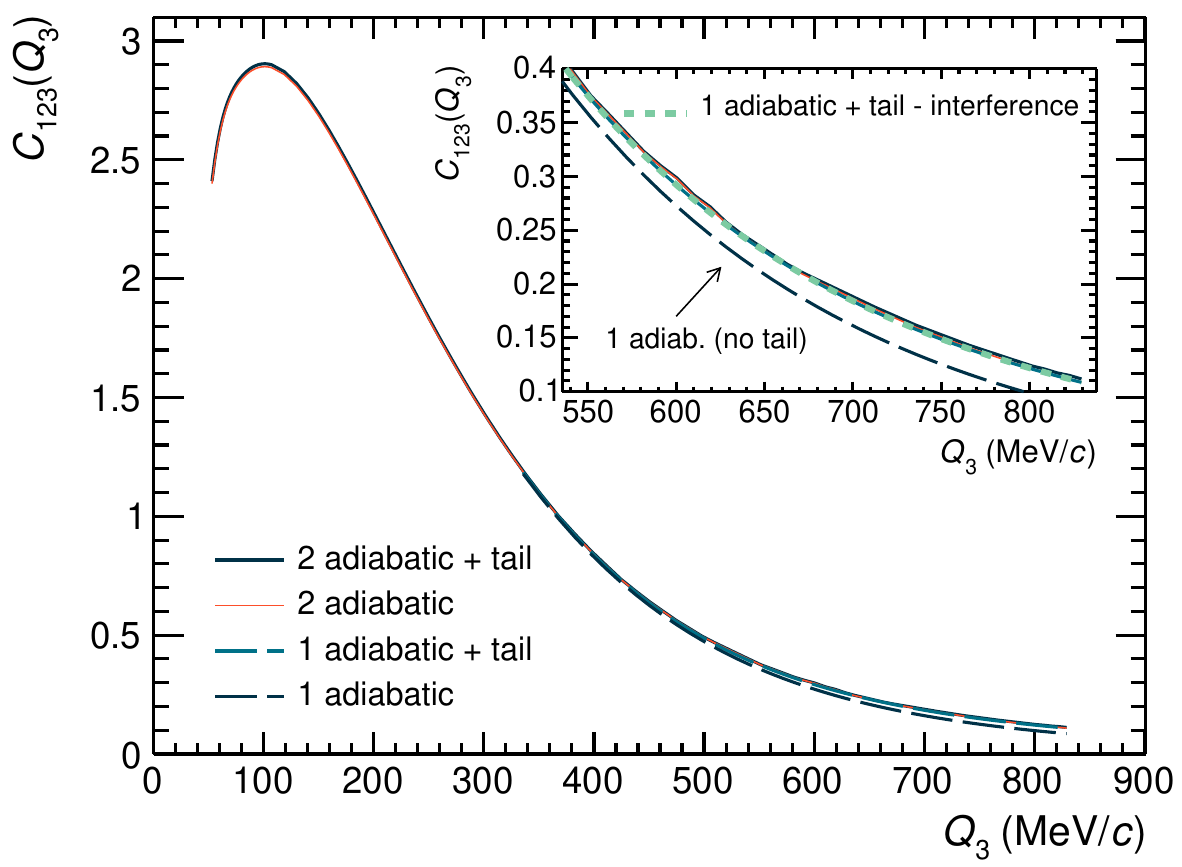}
	\caption{The $J^\pi=1/2^-$ contributions to the correlation function. One or two adiabatic 
	channels have been considered with and without the inclusion of higher contributions treated as free (tail).}
\label{fig:tail}
\end{figure}

The correlation function is 
then computed after inserting the expression given in Eq.(\ref{fullnorm}) into Eq.(\ref{c123}).
It can be seen that the contribution of the interference terms of Eq.(\ref{eq:int}) is very small, since the term 
$K=1$ is not present and
$w^1_{K}(\rho\rightarrow \infty )=0$ for $K>1$, whereas 
$J_{K+2}(Q\rho\rightarrow 0)= (Q\rho/2)^{K+2}/(K+2)!$. 
This is illustrated in Fig.~\ref{fig:tail}, where the contribution to the correlation function arising
from the $J^\pi=1/2^-$ state is shown when one and two adiabatic channels are considered.
The label `tail' indicates that free contributions are included for values of $K>1$ or $K>3$ in each of
the two cases. As we can see, inclusion of the second adiabatic channel slightly modifies the computed
curve only in the large momentum region. In fact, as seen in the inset of the figure (where a zoom
of the curves is shown), the result with one adiabatic channel plus the tail basically overlaps with
the result with two adiabatic channels (for which the inclusion of the tail makes no visible change).
When the square of the wave function is taken, Eq.(\ref{fullnorm}), the interference terms are automatically included. 
However, in the inset of the figure we can also see that removal of the interference contributions 
hardly modifies the computed curve. The conclusion is that one adiabatic channel
is sufficient to treat the interaction, whereas higher channels can be considered as free.

\begin{figure}[t]
\includegraphics[scale=0.44]{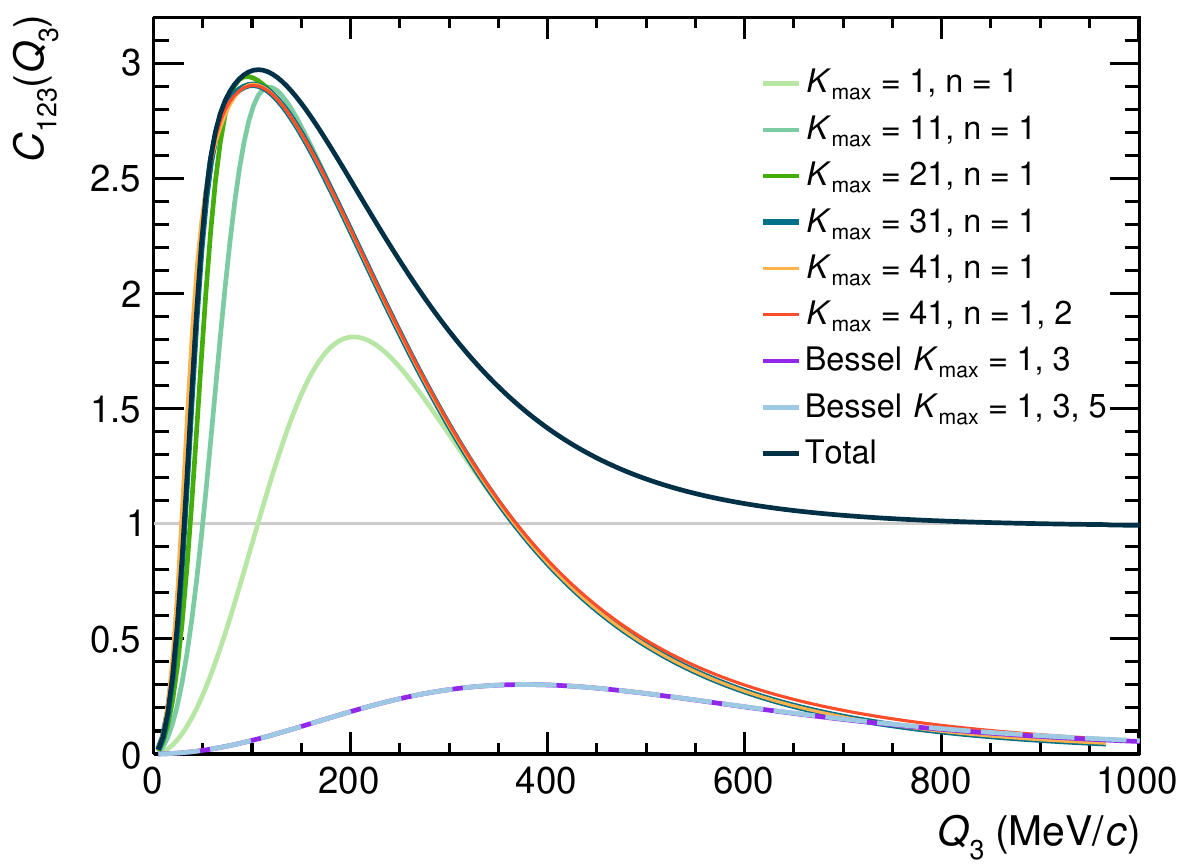}
	\caption{The contribution to the correlation function of the states $J^\pi=1/2^-,3/2^-$ for three neutrons computed with one ($n=1$) or two ($n=2$) adiabatic channels calculated up to the indicated value of $K_\mathrm{max}$. The curves indicated by "Bessel" use Bessel functions as hyperradial functions in the indicated channels (see text). The total curve (black) is the correlation function computed using one adiabatic channel
	for the states $J^\pi=1/2^-,3/2^-$, calculated using HH functions up to $K_{\rm max}=31$, with the other states starting at $K=2$ considered free.}
\label{fig:figcnnn}
\end{figure}

In Fig.~\ref{fig:figcnnn} the correlation function including the $J^\pi=1/2^-$ and $J^\pi=3/2^-$ states with $\rho_0=2\,$fm is analysed. 
The convergence in terms of $K_\mathrm{max}$, the maximum grand-angular momentum quantum number
used to describe the adiabatic potentials, is very fast; a value of $K_\mathrm{max}=31$ 
is already sufficiently accurate. Moreover the lowest adiabatic component, $n=1$, gives
the main contribution to the correlation function. In fact, in the
figure the contribution of the second component, $n=2$, is shown, and
it almost overlaps with the results given just using only one adiabatic component.
The curves labelled by "Bessel $K_{\rm max}=1,3$" and "Bessel $K_{\rm max}=1,3,5$" are calculated
using the Bessel functions as hyperradial function and, as it can be seen, they overlap above 200 MeV with the curves in which the
interaction has been considered. This indicates that
for high values of $Q_3$ the correlation function can be calculated using the free form. The total
curve includes the effects of the interaction in the lowest two adiabatic channels with the rest of the
contribution coming from the other channels considered as free, as given in Eq.(\ref{eq:normf}).

\section{The case of three protons: Including the Coulomb force}

The Coulomb force for a system of three protons is

\begin{equation}
V_\mathrm{Coul}=\sum_{i<j} \frac{e^2}{r_{ij}},
\end{equation}
where $r_{ij}$ is the distance between particles $i$ and $j$. Implementation of this
interaction in a three-body calculation has the significant complication that
the matrix formed by the matrix elements of this potential between different basis 
terms (either within the HH or the HA methods) is not diagonal, even asymptotically.
Furthermore, the asymptotic behaviour of the continuum wave functions is not known 
analytically, and this makes rather difficult to extract the $T$-matrix that enters, 
for instance, in Eq.(\ref{Psihas}).

We postpone a complete treatment of the Coulomb interaction to a forthcoming work,
and explore here the influence of the Coulomb force in the correlation function as an
average of the force on the hyperangles, i.e., 
\begin{eqnarray}
\lefteqn{
V_\mathrm{Coul}(\rho)=\frac{1}{\pi^3}\int d\Omega_\rho \sum_{i<j} \frac{e^2}{r_{ij}} = } \nonumber \\ & &
\frac{3(4\pi)^2}{\pi^3}\int d\alpha \sin^2\alpha \cos^2\alpha \frac{e^2}{\rho\cos\alpha}
=\frac{16}{\pi}\frac{e^2}{\rho},
\label{vcoul}
\end{eqnarray}
which transforms the Coulomb potential into a function depending only on the hyperradius $\rho$. The above equation can be considered as a $0$-term in an expansion of the Coulomb potential in terms of HH multipoles ($K$-expansion). It should be noticed that a symmetric $K=2$ HH function does not exist, and, therefore, the next term of the expansion will be the $4$-term. This property makes operative the hypercentral approximation of the Coulomb in the calculation of the $ppp$ correlation function (for a more complete treatment of the Coulomb interaction see Refs.~\cite{garrido2016,garrido2023}).

When solving the three-body problem, the hypercentral Coulomb potential 
enters directly as a potential in the set of coupled equations providing the hyperradial wave
functions. Disregarding for the moment the nuclear force, the radial equations for 
each value of the grand-angular momentum $K$ are given by:
\begin{equation}
\left( \frac{\partial^2}{\partial z^2} +1 -\frac{2\eta}{z} -\frac{(K+\frac{3}{2})(K+\frac{5}{2})}{z^2}\right) u_K(z)=0\ ,
\end{equation}
with $z=Q\rho$ and $\eta=16 m e^2/(\pi \hbar^2 Q)$, where $m$ is the proton mass.

When $\eta\neq0$ ($ppp$ case), the solution of the equation above is 
\begin{equation}
u_K(z)=F_{K+\frac{3}{2}}(\eta,z),
\label{eq:ukzf}
\end{equation}
which is the regular Coulomb function with order $K+\frac{3}{2}$ and Sommerfeld parameter $\eta$, and which for $\eta=0$ ($nnn$ case) reduces to
\begin{equation}
u_K(z)=zj_{K+\frac{3}{2}}(z)=\sqrt{\frac{\pi z}{2}} J_{K+2}(z).
\label{eq:ukzj}
\end{equation}

Therefore, for the free $ppp$ system, the norm of the continuum wave function is the same as in 
the $nnn$ case, Eq.(\ref{free3n}), but with the replacement:
\begin{equation}
    J_{K+2}(z) \longrightarrow \sqrt{\frac{2}{\pi z}} F_{K+\frac{3}{2}}(z).
    \label{repl}
\end{equation}

 When this is done, we get:
\begin{equation}
|\Psi^0_s|^2_\Omega=\frac{96}{\pi} \frac{1}{(Q\rho)^5} \sum_KF_{K+3/2}^2(Q\rho) N_{ST}(K).
\label{eq:psinorm}
\end{equation}

As in the $nnn$ case, if the antisymmetrization of the three protons is not taken into account,
the number of states is given in Eq.~(\ref{eq:N}), 
whereas in the case of antisymmetrization it is
\begin{equation}
	N_{ST}(K)=N^m_{ST}(K)+4N^a_{ST}(K).
\end{equation}

\begin{figure}[t]
\includegraphics[scale=0.44]{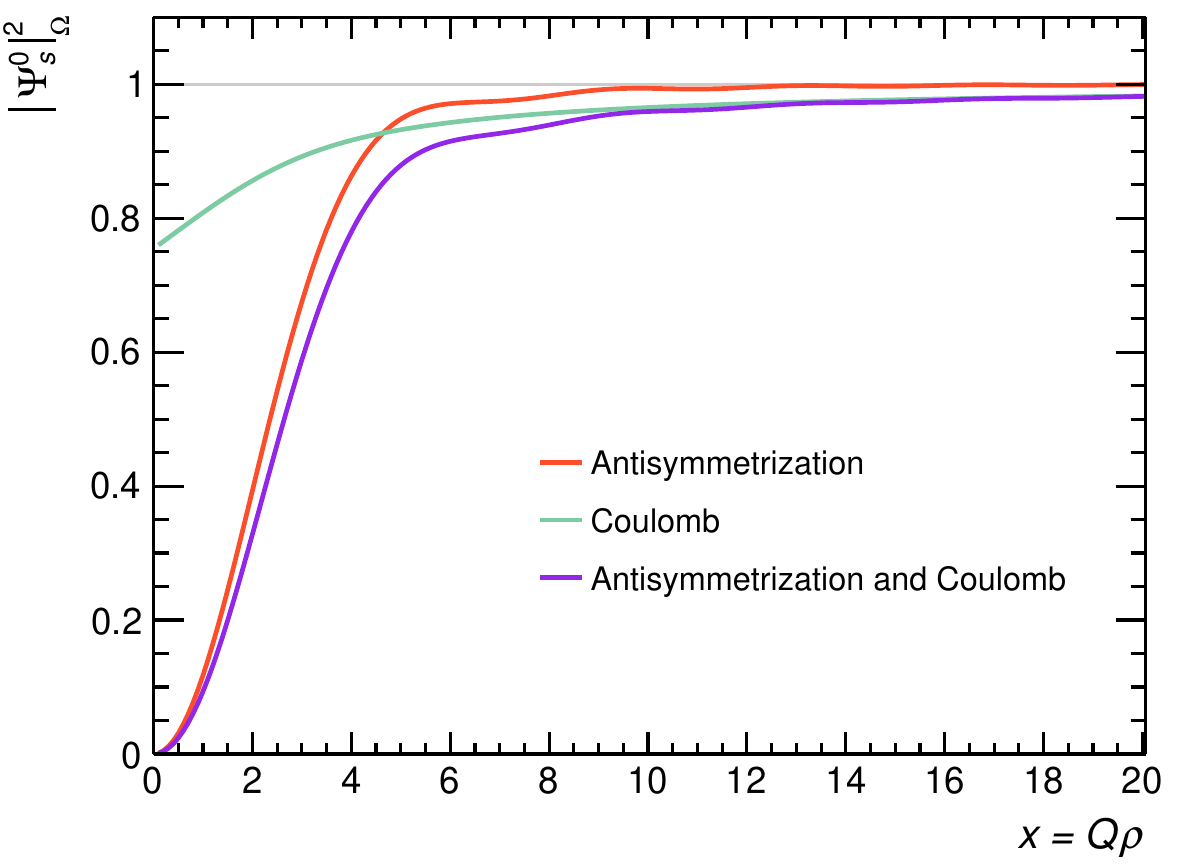}
\caption{The norm of the free scattering wave function for three protons with and without
	considering the hypercentral Coulomb force and antisymmetrization.}
	\label{fig:normp}
\end{figure}

The results for the norm given in Eq.~(\ref{eq:psinorm}) are shown in Fig.\ref{fig:normp}. Note that the curve obtained implementing antisymmetrization, but without the Coulomb interaction, coincides with the one of Fig.\ref{fig3}.

\subsection{Integrating on a spherical source}

The free $ppp$ correlation function obtained after integrating
on a spherical source, Eq.(\ref{c123}), coincides with the one of the $nnn$ system, Eq.(\ref{cnnn}), with the substitution indicated in Eq.(\ref{repl}). We therefore obtain
\begin{equation}
	C_{123}(Q)= \frac{96}{\pi}\frac{1}{Q^5\rho_0^6}\int d\rho\, 
	e^{-\frac{\rho^2}{\rho_0^2}} 
	 \sum_K F^2_{K+3/2}(Q\rho) N_{ST}(K).
\end{equation}

\begin{figure}[t]
\includegraphics[scale=0.44]{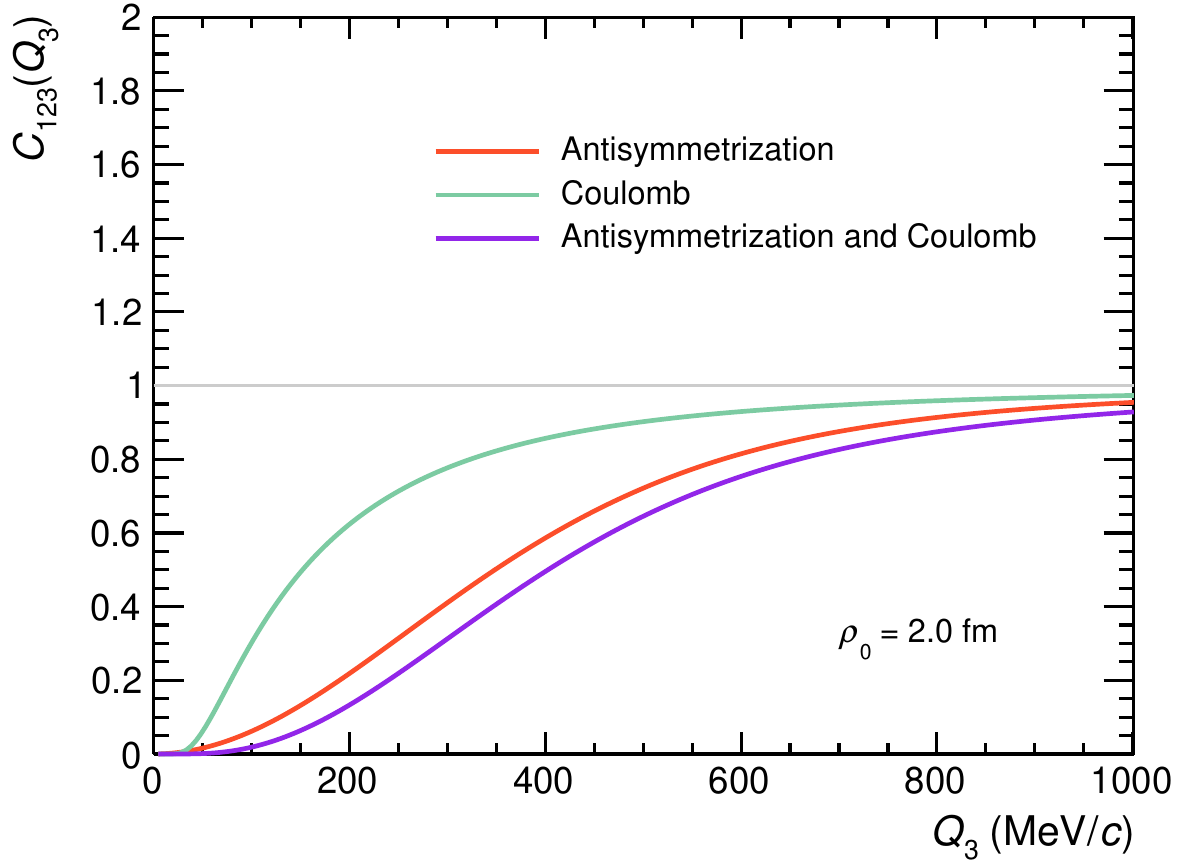}
	\caption{Same as Fig.~\ref{fig:normp} but for the correlation function.}
	\label{fig:corrp0}
\end{figure}

In Fig.\ref{fig:corrp0} we show $C_{123}(Q_3)$ obtained using $\rho_0=2\,$fm, and we compare the no Coulomb case
(red curve) with the Coulomb case (violet curve). The green curve shows the case in which no antisymmetrization
of the three protons is taken into account.

\subsection{Introducing the interaction}

The procedure used to introduce the interaction among the three protons is the same as described for the $nnn$ system. We therefore use the HA expansion method, and the
only difference is that the numerical solutions for the radial functions, $u_n^{n'}$, in Eq.(\ref{3bdad2}), are obtained including
the hypercentral potential of Eq.(\ref{vcoul}).

The consequence is that the asymptotic behaviour of the $u_n^{n'}(Q,\rho)$ functions is now 
\begin{eqnarray}
\lefteqn{
	u_n^{n'}(Q, \rho\rightarrow\infty) \rightarrow 
 } \label{asymc} \\ & &
   i^{K'} \sqrt{\frac{2}{\pi}} 
  \left[\delta_{KK'}F_{K'+\frac{3}{2}}(\eta, Q\rho)+ T_{KK'} {\cal O}_{K'+\frac{3}{2}}(\eta, Q\rho)\right],
  \nonumber
\end{eqnarray}
which is simply the same as for the $nnn$ system, Eq.(\ref{asymu}), but, once more, with the substitution
of Eq.(\ref{repl}), which applies as well for the irregular functions, in such a way that 
${\cal O}_{K}(\eta,Q\rho)=iF_{K}(\eta,Q\rho)+G_{K}(\eta,Q\rho)$.

With the same procedure adopted for the $nnn$ case, we obtain, in analogy with Eq.(\ref{eq:normf}), that the norm of
the scattering wave function for the $ppp$ systems is given by
\begin{eqnarray}
\lefteqn{
	\sum_{J} |\Psi^{J}|^2_\Omega= \frac{96}{\pi}\frac{1}{(Q\rho)^5}  \sum_J (2J+1) \times
    } \label{eq:normfc} \\ & &
  \left(  \sum_{KK'}^{K_0} \left|\frac{u_ K^{K'}(Q,\rho)}{\sqrt{2/\pi}}\right|^2 
	+ \sum_{K>K_0} F_{K+\frac{3}{2}}^2(\eta, Q\rho)  \right) N_{ST}(K)\ ,
	\nonumber
\end{eqnarray}
where, again, $K_0$ is the quantum number indicating the maximum value of $K$ at which the interaction distorts
the free scattering state. Note that the factor $\sqrt{2/\pi}$ that divides the $u_K^{K'}(Q,\rho)$ function is a consequence of
the normalization given in Eq.~(\ref{asymc}), and permits to recover the regular Coulomb function in the free case.

As a first application of the formalism to the $ppp$ case we consider two different models for the $pp$ interaction. Firstly, as in the two-body case,
we model the short-range $pp$ interaction with a Gaussian potential in spin $S=0$, see Eq.(\ref{eq:gausspp}), 
with the parameters selected to reproduce the $pp$
scattering length and effective range of $-7.8063\pm 0.0026$ fm and $2.773\pm0.014$ fm, respectively~\cite{machleidt2011}. In addition we consider also the AV18 NN interaction.

As in the  $nnn$ system, we use the HA method and solve the hyperradial equations for the lowest adiabatic channel.
The only two states having asymptotically the lowest HH channel ($K=1$) compatible with parity 
and antisymmetrization are the $J^\pi=1/2^-$ and $3/2^-$ states. In the same way, the lowest 
adiabatic channel with $K=2$ is consistent with the $J^\pi=1/2^+$, $3/2^+$, and $5/2^+$ states. For the Gaussian interaction, the adiabatic channels calculated with the values $K_{\rm max}=31$ ($K_{\rm max}=30$) odd- (even)-parity states are sufficient to get convergence in the corresponding adiabatic channels. However when using the AV18 interaction much higher values are considered, $K_{\rm max}=151$ ($K_{\rm max}=150$) for odd- (even)-parity states.
After solving the adiabatic equations to obtain the hyperradial functions $u_K^{K'}(Q,\rho)$, the norm of the scattering $ppp$ wave function is completely determined. The states with $J^\pi=1/2^-,3/2^-$ and
$J^\pi=1/2^+,3/2^+,5/2^+$ are given by the first term of Eq.(\ref{eq:normfc}) with $K_0=2$, whereas the second term includes all the other channels.
Making use of the norm, and integrating over $\rho$ as given in Eq.(\ref{c123}), we
obtain the $ppp$ correlation function for each of the cases.

\begin{figure}[t]
\includegraphics[scale=0.44]{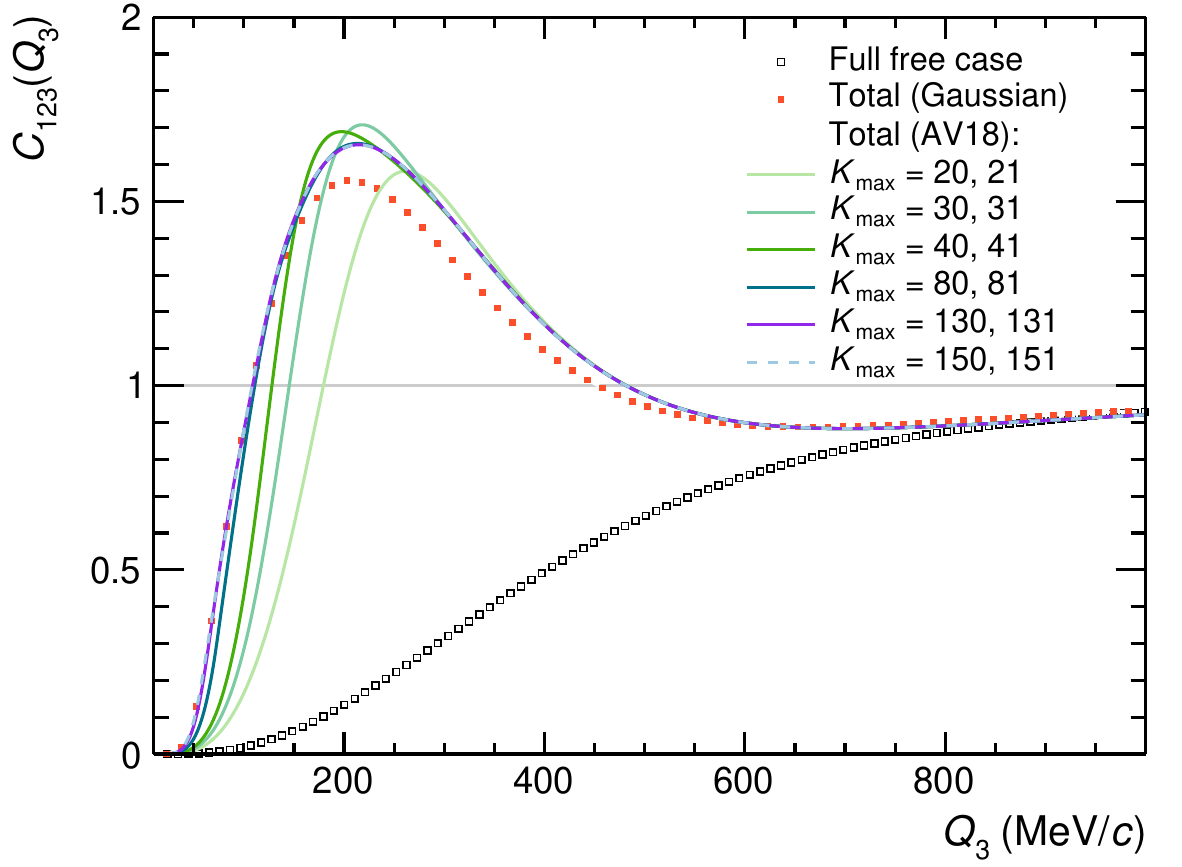}
	\caption{The $ppp$ correlation function with a source sizes of $\rho_0=2\,$fm, calculated using the
	first adiabatic channel for asymptotic states with $K=1,2$, considering different values of $K_{\rm max}$. For $K>2$ the free scattering function is considered. The full free case is also shown. }
	\label{fig:cpppKmax}
\end{figure}

We first study the impact of $K_{\rm max}$ on the correlation function. In Fig.\ref{fig:cpppKmax} this is shown for the two potential models and using a source size value of $\rho_0$=$2\,$fm. As mentioned, the interaction has been included in the $J^\pi=1/2^-,3/2^-$ states
and $J^\pi=1/2^+,3/2^+,5/2^+$ states with $K_0=1$ and $K_0=2$, respectively. The free scattering wave function has been considered for $K>2$. In other words, the figure shows the correlation function obtained using Eq.(\ref{eq:normfc}) with $K_0=2$ and, for the AV18 potential, at different values of $K_{\rm max}$ in the computation of the hyperradial functions $u_K^{K'}(Q,\rho)$. It is interesting to observe that 
the converged results for the Gaussian and the AV18 potential are on top of each other in the very low $Q$ region. However the
values of $K_{\rm max}$ needed to reach convergence are very different. In fact, as can be seen in the figure, the AV18 curves for $K_{\rm max}\ge 130$ almost overlap. Moreover, we can observe a higher peak when this potential is used. This is due to contributions beyond the pure $s$-wave at which the gaussian potential is limited. We also see that increasing $Q$ the effect of the short-range interaction smears out and the correlation function is well described by the free scattering wave function.

\begin{figure}[t]
	\includegraphics[scale=0.44]{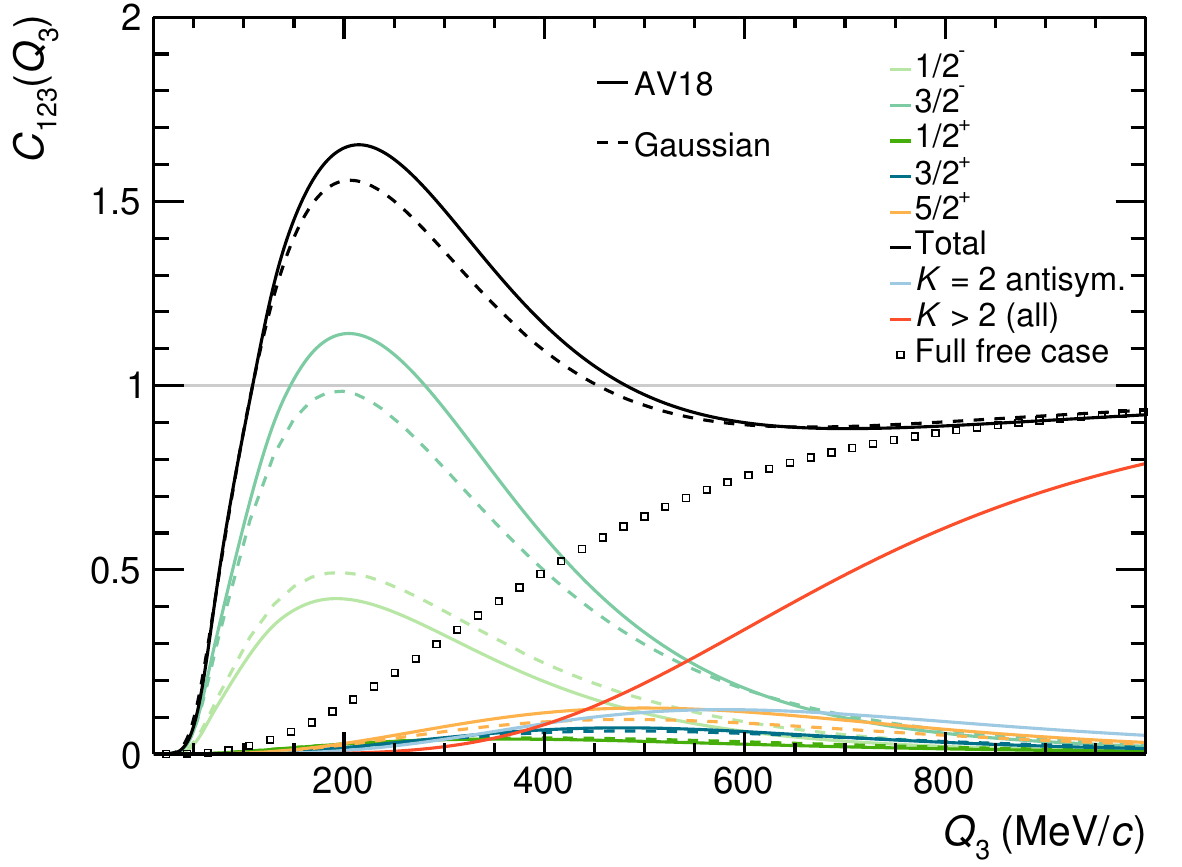}
	\caption{The $ppp$ correlation function using the AV18 potential (solid lines) and compare to the results using a Gaussian interaction (dashed lines). The source size $\rho_0$=2.0 fm has been used. The Coulomb free contribution is also shown.}
	\label{fig:pppav18}
\end{figure}

We now study different partial wave contributions to the $ppp$ correlation function. This is shown in Fig.~\ref{fig:pppav18} where the different $J^\pi$-contributions are shown explicitly for the two potential models. The results have been computed using a source size of $2.0\,$fm. We observe that the use of the realistic force slightly increases the observable value around the peak, which is mainly due to the contribution of the $J^\pi=3/2^-$ state. As evident from the figure, the peak of the observable is almost completely constructed by the odd-parity components to the wave function. This means that detailed measurements of the $ppp$ correlation function around the peak could be used to asses the capability of the potential models to produce the correct splitting of the three-body $P$ waves phase-shifts. A problem already observed in the description of asymmetries as the $pd$ $A_y$ analyzing power (for a recent discussion see Ref.~\cite{girlanda2023}). 

To complete the study of the correlation function, we have also performed calculations using the AV18 potential plus the Urbana IX three-body force. We have observed differences, in the direction of reducing the correlation function, of the order of 1\% or lower. This confirms the very small effect of the three-body force in the $ppp$ and
$nnn$ systems already observed in Ref.\cite{higgins2020} for the latter. This is mainly due to the Pauli principle which prevents the three equal nucleons to be close enough to feel the influence of the three-nucleon force.

\begin{figure}[t]
\includegraphics[width=0.48\textwidth]{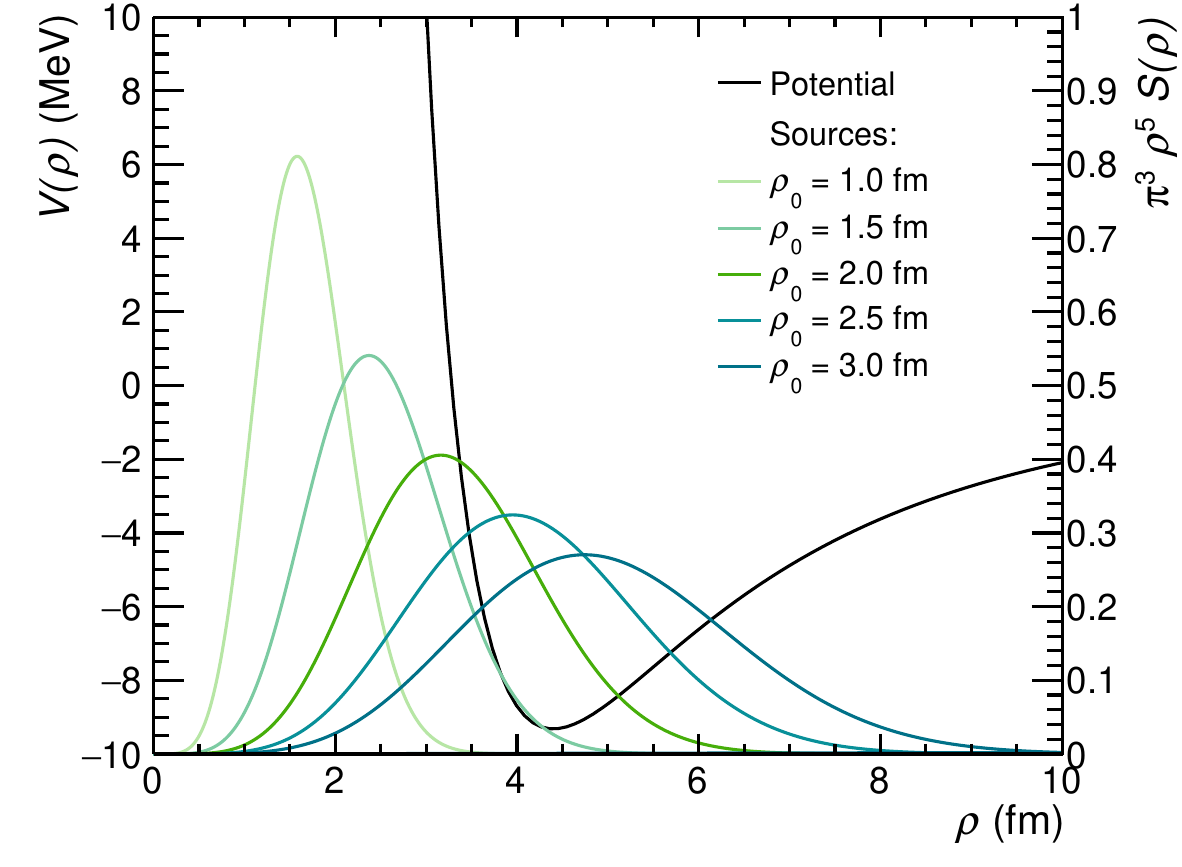}\\
\includegraphics[width=0.48\textwidth]{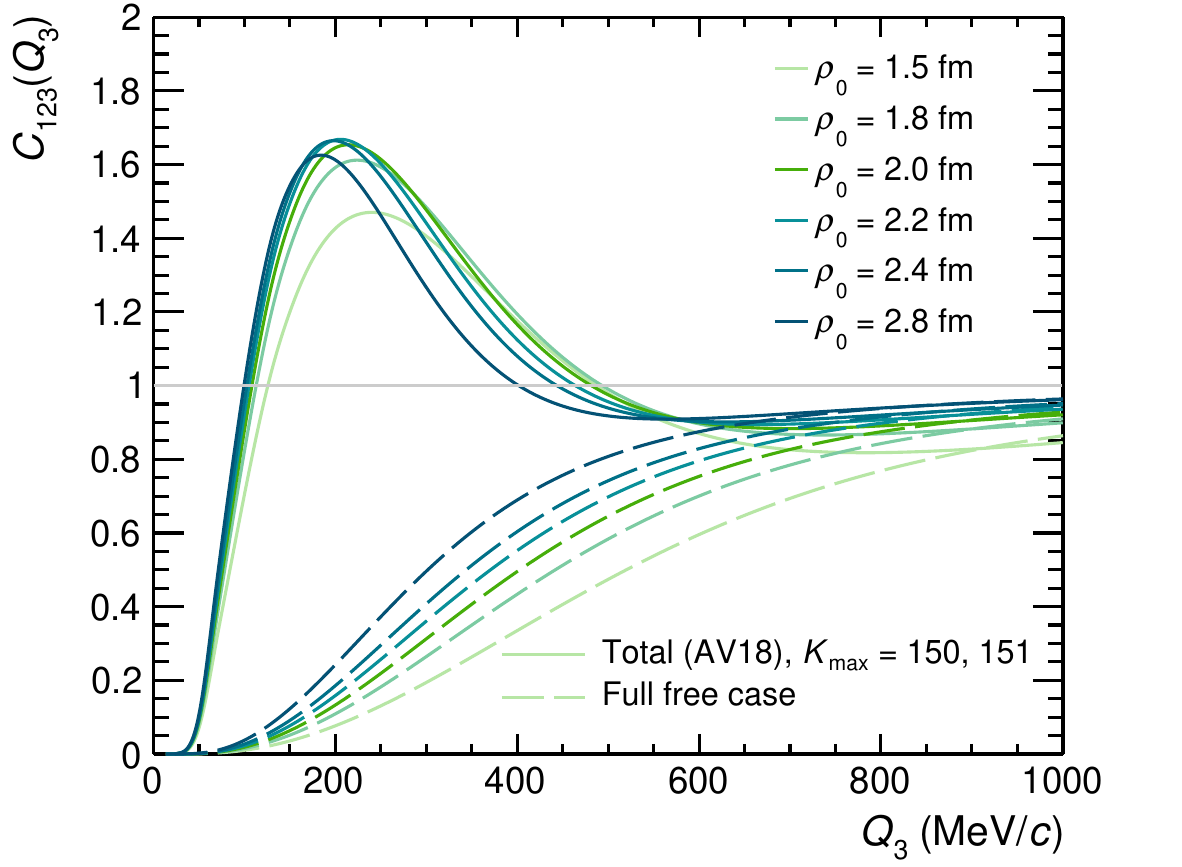}
\caption{{\it Top panel:} Overlap between the average $s$-wave potential and the source function for different radii. \\
{\it Bottom panel:} The $ppp$ correlation function calculated using the AV18 potential for different source radii (solid lines). The Coulomb free contribution are shown with dashed lines.}
\label{fig:v18r}
\end{figure}

Finally we discuss the effects of the source size. In the top panel of Fig.~\ref{fig:v18r}, we show, as a function of the hyperradius, the overlap of the source function,
$\pi^3\rho^5 S_{123}(\rho)$, with the short-range potential averaged on the hyperangles, i.e. defined as
\begin{equation}
V(\rho)=\frac{1}{\pi^3}\int d\Omega_\rho \sum_{i<j}V_0(r_{ij}) \ .
\end{equation}
For this purpose we have used the $s$-wave part of the AV18 potential.
As can be seen from the figure, the effect of the potential is more appreciable for source sizes larger than $2\,$fm. In the bottom panel, the correlation function is shown for the different source radii. Moreover, the
correlation function considering only the Coulomb potential is explicitly shown as dashed lines.

\subsection{Comparison with the experimental data}
\label{subsec:comparison}

\begin{figure}[t]
	\includegraphics[scale=0.44]{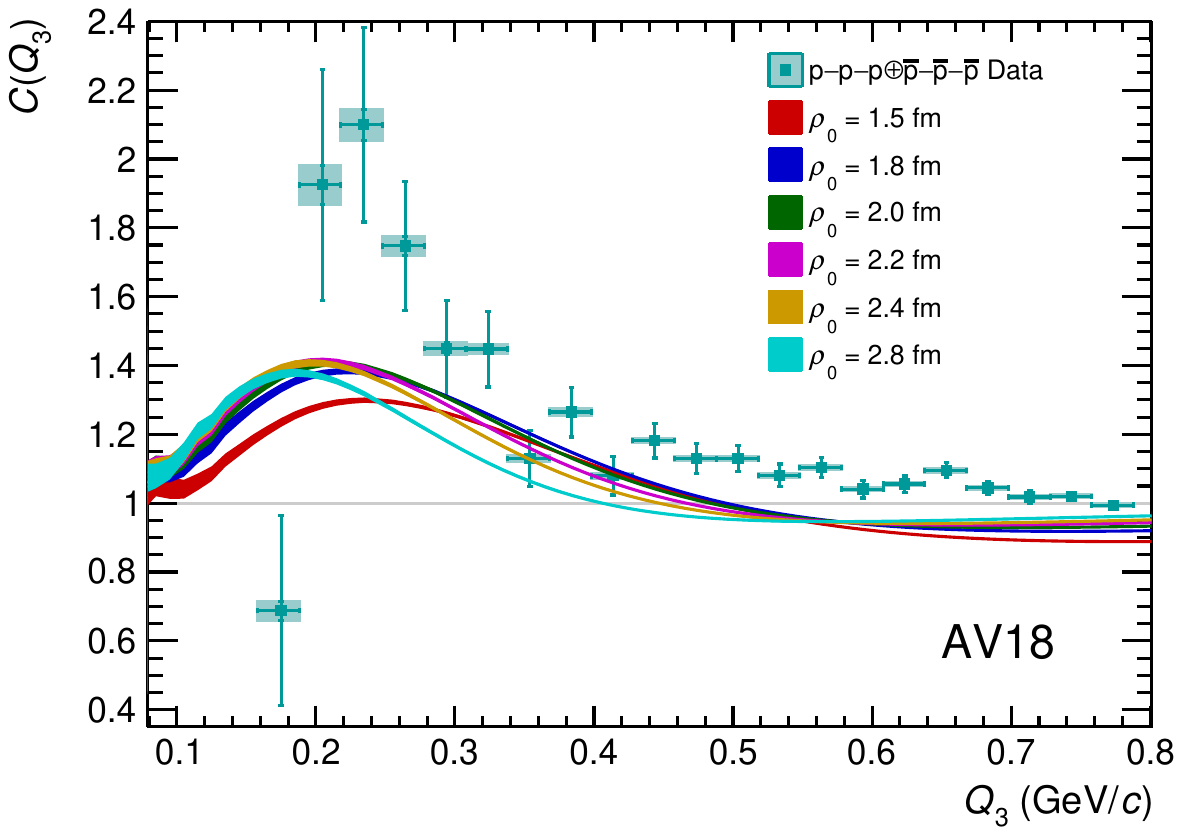}
	\caption{The comparison of the $ppp$ correlation function measured by the
		ALICE Collaboration~\cite{femtoppp} (cyan full squares) and the
		calculated correlation functions corrected for experimental
		effects (bands). The vertical lines on the data points
		correspond to the statistical uncertainties and the boxes to the
		systematic ones. Different colour bands correspond to the
		different source sizes. For details of corrected correlation
		functions, we refer to the text.}
	\label{fig:comparePPPtoData}
\end{figure}

In this Section, the calculated $ppp$ correlation function is compared to the experimental one published by the ALICE Collaboration in Ref.~\cite{femtoppp}. The experimental correlation function is shown in Fig.~\ref{fig:comparePPPtoData} as cyan squares~\footnote{The first bin, published by the ALICE Collaboration, is not shown here, however it is at correlation function value of 3.5.}. The vertical lines correspond to the statistical uncertainties while the boxes denote systematic ones. The measured three-body correlation function is obtained in a similar way as explained for the two-body case in Section~\ref{subsec:comparison2b}. However, Eq.~(\ref{eq:CFdefinition2}) is extended to three particles as
\begin{equation}
    C(Q_3) = \mathcal{N}\frac{A(Q_3)}{B(Q_3)} \ .
    \label{eq:CFdefinition3}
\end{equation}
Here, $A(Q_3)$ corresponds to the same event triplets, while the distribution $B(Q_3)$ is obtained by combining three particles from three different events. The observable $Q_3$ is Lorentz invariant and is estimated experimentally as $Q_3 = \sqrt{-q_{12}^{2} - q_{23}^2 - q_{31}^2}$, where $q_{ij}$ is the norm of the four-vector corresponding to the relative momentum between particles $i$ and $j$~\cite{femtoppp}. The normalization constant $\mathcal{N}$ is such that the measured correlation function is equal to unity at $Q_3$ region, where particles are expected to not interact anymore via final state interaction~\cite{femtoppp}. 

As has been explained previously, the calculated correlation functions cannot be compared directly to the measured ones since the experimental effects such as momentum resolution and the presence of secondary and misidentified protons must be taken into account. The momentum resolution can be included by performing a convolution of the theoretical correlation function, the mixed event distribution and the momentum resolution matrix as described in Ref.~\cite{pLambda}. This has been done by employing the mixed event distribution and momentum resolution matrix recently published by the ALICE Collaboration in Ref.~\cite{publicNote}. The effect of secondary and misidentified protons can be taken into account by employing Eq.~(\ref{eq:2Bcorrection}) extended to the three-particle case as
\begin{equation}
    C (Q_3) = \lambda_{ppp} C_{ppp} (Q_3) + \lambda_{ppp_\Lambda} C_{ppp_\Lambda} (Q_3) + \lambda_\mathrm{X} C_\mathrm{X} (Q_3) \ .
\label{eq:3Bcorrection}
\end{equation}

$C_{ppp} (Q_3)$ is here included as the calculated correlation functions shown in Fig.~\ref{fig:v18r} with the fraction of genuine $ppp$ pairs being $\lambda_{ppp}$ = 0.618~\cite{femtoppp}. The residual $pp\Lambda$ correlation function 
$C_{ppp_\Lambda} (Q_3)$, where one of the protons is a product of $\Lambda$
decay, is obtained by taking $C_{pp\Lambda} (Q_3)$ correlation function and
performing a convolution with the decay matrix which maps the $Q_3$ of initial
$pp\Lambda$ triplet to the $Q_3$ of resulting $ppp$ triplet after the $\Lambda$
decay. The $pp\Lambda$ correlation function is known only experimentally and has
been measured by the ALICE Collaboration~\cite{femtoppp}. However, the published
values are provided only up to $Q_3$ = 0.8 GeV/$c$. When used to estimate
$C_{ppp_\Lambda} (Q_3)$ correlation function, this results in edge effects at
the range $Q_3>0.55$ GeV/c. For this reason, the $pp\Lambda$ correlation
function at larger values than $Q_3$ = 0.8 GeV/$c$ is assumed to be equal to the
projector values published in Ref.~\cite{femtoppp} which is shown to be in good
agreement with data.  The fraction of $pp\Lambda$ triplets in the experimental
sample is $\lambda_{pp\Lambda}$ = 0.196~\cite{femtoppp}. In
Eq.~(\ref{eq:3Bcorrection}), the $\lambda_\mathrm{X} C_\mathrm{X} (Q_3)$ term
includes the rest of contributions, such as protons stemming from other hyperon
decays or misidentified proton contribution. The correlation functions for
$C_{ppH} (Q_3)$, where $H$ denotes other hyperons than $\Lambda$ which decay to
proton, are not known. However, the $\lambda$ parameters for single channels are
very small and thus, after accounting for the decay kinematics, $C_{ppp_H}
(Q_3)$ are expected to be equal to unity. The correlation function for
misidentified protons is also assumed to be equal to unity. Thus $C_\mathrm{X}
(Q_3)$ in Eq.~(\ref{eq:3Bcorrection}) is also assumed to be equal to unity. The
resulting correlation functions $C(Q_3)$ are shown in
Fig.~\ref{fig:comparePPPtoData} as colourful bands for different source sizes
already accounting for momentum resolution effects. The bands include propagated
systematic and statistical uncertainties of  $C_{pp\Lambda} (Q_3)$.

\textcolor{blue}{The comparison in Fig. 12 shows that the calculated correlation functions are systematically below the experimental data in the entire range $Q_3 < 0.8$ GeV/$c$, independently on the chosen value for the source size $\rho_0$. On the contrary, the shape is well reproduced. The experimental correlation function obtained in Ref.~\cite{femtoppp} was normalised to the unity in the range $1.0 < Q_3 < 1.2$ GeV/$c$, assuming that any effect induced by the final state interaction of the three protons is absent in such an interval. This argument was motivated by the analysis of the $pp$ pair correlations in $ppp$ triplets and Monte Carlo studies (see Ref.~\cite{femtoppp} for more details). Interestingly, the calculation performed in this paper evidences the presence of residual correlations also at large $Q_3$. These correlations result in a dominantly repulsive effect at $Q_3 >$ 0.5 GeV/$c$ mainly due to the antisymmetrisation of the $ppp$ wave function. A proper comparison between the data and the theory would require extending the experimental measurements to a larger $Q_3$ interval, including the region where the theoretical correlation function converges to unity. To be noticed that this study has to be considered preliminary as further improvements in the modelling are required concerning, in particular, the source function. }

\section{Conclusions}

In the present work we have made a detailed analysis of the $nnn$ and $ppp$
correlation functions. This study was motivated by the recent effort to measure
the $ppp$ correlation function from high-energy $pp$ collisions at the LHC.
The main difficulties in the computation of this observable arise from the 
complicate structure of the asymptotic $ppp$ wave function induced by
the long range Coulomb interaction and the different spatial-spin structures 
needed to fulfill the Pauli principle requirements. In this analysis we performed a
simplification and treated the Coulomb interaction using the hypercentral
approximation. This approximation is well motivated by the absence of the $K=2$ term in the expansion of the Coulomb interaction in terms of HH functions. Accordingly we can solve the associate dynamical equations
with boundary conditions similarly to the $nnn$ case. To this end  we made use
of HA method showing that the lowest adiabatic potential already gives a
extremely accurate description of the dynamics. In a first step we have used a simple potential model consisting in a Gaussian interaction parametrized in order to reproduce the $pp$ scattering length and effective range. This simplified interaction captures most of the structure of the observable in both, the two-body and the three-body sectors. This is mainly motivated by the large value of $pp$ scattering length when the Coulomb interaction is switched off. Motivated by the encouraging results in the description of the $ppp$ correlation function using the mentioned approximations, in a second step we consider the AV18 interaction. The main modification with the precedent results was found in the $3/2^-$ state with a higher peak and, consequently an overall increase of the observable. To be noticed that the inclusion of the Urbana IX three-body force, very important in the description of nuclei, gives a negligible contribution in the correlation function mainly due to the low probability of having three protons close to each other.

The present results complete the study of the three-nucleon correlation function
initiated with the study of the $pd$ correlation function~\cite{femtopd,pdtheory}. It
shows that the measurements of these observables from high energy collisions open the door to a new way of studying reactions in different three-body systems. In particular it would be possible to use the present method, the HH basis in conjunction with the HA expansion, to describe the $pp\Lambda$ correlation function already measured by the ALICE Collaboration.

\acknowledgments 
The authors gratefully acknowledge Laura Fabbietti for the fruitful discussions that helped the improvement of the paper.   
This work has been partially supported by: Grant PID2022-136992NB-I00 funded by MCIN/AEI/10.13039/501100011033 and, as appropriate, by “ERDF A way of making Europe”, by the “European Union” or by the “European Union NextGenerationEU/PRTR”; the Deutsche Forschungsgemeinschaft through Grant SFB 1258 “Neutrinos and Dark Matter in Astro- and Particle Physics”; the Deutsche Forschungsgemeinschaft (DFG, German Research Foundation) under Germany's Excellence Strategy – EXC 2094 – 390783311.

\appendix
\section{The source for two and three particles}   
\label{appendixA}

The correlation function of a pair of (identical) particles
can be written in general as (here, for simplicity, we disregard the particle spins)
\begin{eqnarray}
 C_{12}\left(\vecp_1, \vecp_2\right)
 &=&  \int d^3 r_1\,d^3 r_2 S_1
 \left(r_1\right) S_1\left(r_2\right)\nonumber \\
 &&\times \abs{\Psi(\vecp_1,\vecp_2,\vecr_1,\vecr_2)}^2\ ,
\label{eq:2pp}
\end{eqnarray}
where $\Psi(\vecp_1,\vecp_2,\vecr_1,\vecr_2)$ denotes the
two-particle scattering wave function that asymptotically describes
particle 1 (2) with momentum $\vecp_1$ ($\vecp_2$).
In Eq.~\eqref{eq:2pp} $S_1(r)$ describes the spatial shape of the source for
single-particle emissions.
It can be approximated as a Gaussian probability distribution with a width
$R_M$, which is defined as follows:
\begin{equation}
  S_1(r) = \frac{1}{(2\pi R_M^2)^{\frac32}} e^{{-}r^2/2 R_M^2}\ .
\label{eq:S1}
\end{equation}
$R_M$ is also known as the source size for single particle emission.
Eq.~\eqref{eq:2pp} can be simplified by noting that in the wave
functions the dependence on the overall center-of-mass (CM) coordinate
can be trivially factored out.
Introducing the CM coordinate $\bm{R}_{CM} \equiv \frac{\vecr_1+\vecr_2}{2}$ and the relative distance $\vecr \equiv \vecr_1-\vecr_2$,
and rewriting the two-particle wave function as
$\Psi(\vecp_1,\vecp_2,\vecr_1,\vecr_2)= e^{{-}i\bm{R}_{CM}\cdot\bm{P}}\psi_{\veck}(\vecr)$ leads to
the Koonin-Pratt relation for two-particle correlation function~\cite{Koonin,Pratt},
which we write here as
\begin{equation}
 C_{12}(k) = \int d^3 r \, S(r)
 \abs{\psi_{\veck}\left(\vecr\right)}^2 \ ,
\end{equation}
where $\psi_{\veck}\left(\vecr\right)$ represents the
two-particle relative wave function, with $\veck=(\vecp_1-\vecp_2)/2$, and $S(r)$ is the two-particle emission source, given by
\begin{equation}
 S(r) = \left(\frac{1}{4 \pi R_M^2}\right)^{\!3/2}
 e^{{-}\frac{r^2}{4 R_M^2}} \ .
\label{eq:twoParticleSource}
\end{equation}
This source is identical to that given in Eq.~(\ref{sour2b}), so we can identify $R$ with the single-particle source radius $R_M$. 

The generalization to the calculation of the correlation function of three identical particles is given by 
\begin{equation}\label{eq:c123}
 C_{123}(Q) = 
 \int d^3r_1 d^3r_2 d^3r_3\; S_1(r_1) S_1(r_2) S_1(r_3) |\Psi_s|^2\ ,
\end{equation}
where $Q$ and $\Psi_s$ have been defined at the beginning of Section~\ref{chap:threebodycase}. Since we are considering the case of three identical particles,
for all of them the source is assumed to have the same form as that given in Eq.~(\ref{eq:S1}).

Eq.~\eqref{eq:c123} can be simplified by introducing the following variables:
\begin{equation}
\bm{x} =  \bm{r}_2-\bm{r}_2\ ,
 \quad \bm{y}'=\bm{r}_3-\frac{\bm{r}_1+\bm{r}_2}{2}\ ,
 \quad \bm{R}_3=\frac13(\bm{r}_1+\bm{r}_2+\bm{r}_3)\ .
\end{equation}
Now
\begin{equation}
 d^3r_1 d^3r_2 d^3r_3 = d^3R_3 d^3x d^3y'\ ,
\end{equation}
and
\begin{equation}
 S_1(r_1) S_1(r_2) S_1(r_3)
 = \frac{e^{-(3R_3^2+{\frac23}y^{\prime 2}+{\frac12}x^2)/2R_M^2}}%
   {(2\pi R_M^2)^{\frac92}} \ .
\end{equation}
Integrating over $d^3R_3$ (the wave function $\Psi_s$ does not depend on $R_3$), we obtain
\begin{equation}
 C_{123}(Q) =  \int d^3x d^3y'\; \frac{e^{-({\frac43}y^{\prime 2}+x^2)/4R_M^2}}
 {(3\pi R_M^2)^{\frac32}(4\pi R_M^2)^{\frac32}} \abs{\Psi_s}^2 \ .
\label{eq:mrow7}
\end{equation}
In term of the Jacobi vector $\bm{y}=\sqrt{\frac43}\bm{y}'$ (see the beginning of Section~\ref{chap:threebodycase}), this integral can be rewritten as
\begin{multline}
 C_{123}(Q) =  \left({\frac34}\right)^{\frac32}
 \int d^3x d^3y \\
 \null \times \frac{e^{{-}(x^2+y^2)/4R_M^2}}{(3\pi R_M^2)^{\frac32} (4\pi R_M^2)^{\frac32}}
 \abs{\Psi_s}^2  \ .
\label{eq:mrow8}
\end{multline}
Introducing (see again the beginning of Section~\ref{chap:threebodycase}) the  hyperradius, given by $\rho=\sqrt{x^2+y^2}$,
and the hyperangular variables $\Omega_\rho$~\cite{HH2008}, such that
$ d^3x d^3y=\rho^5 d\rho d\Omega_\rho$, we finally obtain
\begin{equation}
 C_{123}(Q) =   \int \rho^5 d\rho d\Omega_\rho\;
 S_{123}(\rho)
 \abs{\Psi_s}^2 \ ,
\label{eq:mrow9}
\end{equation}
where
\begin{equation}
 S_{123}(\rho) =  
 \frac{e^{{-}\rho^2/4R_M^2}}{(4\pi R_M^2)^3} \ ,
\label{eq:s123}
\end{equation}
and which after comparison with Eqs.~(\ref{c123}) and~(\ref{sour3b}) permit to identify $\rho_0=2 R_M$ (see however the discussion at the end of Section V.C). 

Starting with a Gaussian shape for the single-particle emission, the two- and three-particle emission sources result spherical. Accordingly,
the correlation functions in Eq.~(\ref{corr}) and Eq.~(\ref{c123}) can be 
obtained after replacing $|\Psi_s|^2$ by the "angle averaged" square modulus of the wave functions. 

For example, in the experimental detection of two protons, all pairs with the same $k$ but arbitrary direction $\Omega_k$ are counted to construct $C_{12}(k)$. This is equivalent to perform the average $\frac{1}{4\pi}\int d\Omega_k |\Psi_s|^2$, and the correlation function in Eq.~(\ref{corr}) can then be computed as:
\begin{equation}
 C_{12}(k)=\int r^2 dr d\Omega_r S_{12}(r) 
\left[  \frac{1}{4\pi} \int d\Omega_k |\Psi_s|^2 \right].
\label{eqa13}
\end{equation}

Since the source function is spherical, Eq.~(\ref{eqa13}) can be trivially rewritten as:
\begin{equation}
 C_{12}(k)=\int r^2 dr d\Omega^\prime_r S_{12}(r) 
\left[  \frac{1}{(4\pi)^2} \int d\Omega_r \int d\Omega_k |\Psi_s|^2 \right],
\label{eqa14}
\end{equation}
which makes clear that the correlation function $C_{12}(k)$ can be computed as given in 
Eq.~(\ref{corr}), but replacing $|\Psi_s|^2$ by the "angle averaged" norm of the scattering wave function as defined in Eq.(\ref{norm}).

The same situation results in the three body case. The detection of three protons at a fixed value of $Q$ is equivalent to perform the average over $\Omega_Q$. Noting
that $\int d\Omega_Q=\int d\Omega_\rho=\pi^3$, and following the same reasoning as
shown above for the two-body case, we have that the correlation function $C_{123}(Q)$ can also be computed as given in Eq.(\ref{c123}), but replacing now
$|\Psi_s|^2$ by the "angle averaged" norm of the scattering wave function given in Eq.(\ref{norm3b}).

\end{document}